\documentclass[11pt]{JHEP}
\usepackage{latexsym}
\usepackage{amssymb,amsfonts}
\font\mybb=msbm10 at 10pt
\def\bb#1{\hbox{\mybb#1}}
 \preprint{
July 31, 2012. V2: August 28, 2012}
\renewcommand{\theequation}{\arabic{section}.\arabic{equation}}
\title{On pure spinor formalism for quantum superstring and spinor moving frame}

\author{
Igor A. Bandos
  \\   Department of
Theoretical Physics, University of the Basque Country
 \\ UPV/EHU,
P.O. Box 644, 48080 Bilbao, Spain
\\ and \\ IKERBASQUE, Basque Foundation for Science, 48011, Bilbao, Spain}

\date{
July 31, 2012; V2: August 28, 2012, printed \today }

\abstract{The D=10 pure spinor constraint can be solved in terms of spinor moving frame variables $v^{-\alpha}_q$ and  8-component   complex null vector $\Lambda^+_q$,   $\Lambda^+_q\Lambda^+_q=0$, which can be related to the $\kappa$--symmetry ghost. Using this and similar solutions for the conjugate pure spinor and other elements of the non-minimal pure spinor formalism we present a (hopefully useful)  reformulation of  the  measure of the pure spinor path integral for superstring in terms of products of Cartan form corresponding to the coset of 10D Lorentz group and to the coset of complex orthogonal group $SO(8,C)$.
Our study suggests a possible complete reformulation of the pure spinor superstring in terms of new irreducible set of variable.

}


\begin{document}

\section{Introduction}

The pure spinor approach \cite{Berkovits:2000fe} is very successful in the description of quantum superstring. It provides the way of covariant calculation of loop superstring amplitudes \cite{Berkovits:2004px,Berkovits:2005bt,Berkovits:2006vi} (see  \cite{Gomez:2010ad,Mafra:2012kh} for recent progress and more references).  However, despite certain progress reached in \cite{Matone:2002ft}  and \cite{Berkovits:2004tw}, its origin and relation with classical Green--Schwartz action is still to be clarified more. Furthermore, the origin and structure  of the path integral measure
\cite{Berkovits:2006vi,Berkovits:2006ik,Oda:2007ak}, which is used to obtain the superstring loop amplitudes in the frame of pure spinor formalism, remains mysterious.

The recent study in  \cite{Grassi:2011ie} addressed this problem by trying to extract the pure spinor measure $d^{11}\lambda$ from the ${\bb C}^{16}-\{0\}$ integration measure characteristic for the space of 10D Weyl spinors.

In this paper we develop an approach which can be considered as 'bottom--up' with respect to \cite{Grassi:2011ie}. To begin with, instead of trying to extract the pure spinor measure  from the measure on a bigger space, we decompose $d^{11}\lambda$  on two factors which look less mysterious:  one of these is the measure on the space ${\bb C}^7-\{ 0\}$
of 8-dimensional complex null-vectors $\Lambda^+_q$ (described, in particular, in  Appendix of \cite{Grassi:2011ie}) and second is constructed
from  four vielbein forms of ${\bb S}^8$ space (which can be identified with a celestial sphere of 10D observer)\footnote{The complex 11 degrees of freedom in 10D pure spinor $\lambda^\alpha$ are reproduced in our decomposition as 7+4 where 7=8-1 is the dimension of the space ${\bb C}^7-\{ 0\}$ spanned by the complex null--vector $\Lambda^+_q$, obeying  $\Lambda^+_q{\Lambda}{}^+_q=0$, while 4=8/2 corresponds to 8 real dimensions of ${\bb S}^8$.} with the use of the $SO(8)$ Klebsh--Gordan coefficients $\gamma^i_{q\dot{p}}$, the above complex null-vectors $\Lambda^+_q$ and its complementary ${\Lambda}{}^-_q$ (another complex null vector which obeys $\Lambda^+_q{\Lambda}{}^-_q=1$) \footnote{These two are complex and are used to obtain a kind of holomorphic 4--form spin tensor on ${\bb S}^8$ space (of 8 real dimensions) which contributes to the holomorphic 11-form $d^{11}\lambda$. Notice that this does not imply an introduction of complex structure on just ${\bb S}^8$ manifold; rather it suggests a possible  complex structure on the space of pure spinor considered as 11 dimensional fiber bundle over ${\bb C}^7-\{ 0\}$ with the fiber ${\bb S}^8$.}.

After finding the above described expression for the pure spinor  $d^{11}\lambda$, we apply the similar method to the other elements of the path integral measure of the non--minimal pure spinor formalism, beginning from  $d^{11}\bar{\lambda}$ for 'conjugate spinor' $\bar{\lambda}_\alpha$ which obeys, besides the pure spinor constraints, the conditions ${\lambda}^\alpha\bar{\lambda}_\alpha\not=0$. Already at this stage it is useful to complete the set of two complex 8 component null vectors  $\Lambda^\pm_q$ till the $SO(8,\bb{C})$ valued matrix. We use this '$SO(8,\bb{C})$ frame' matrix field to solve the constraints on the  fermionic spinors $r_\alpha$ of the non-minimal pure spinor  formalism. However, this solution includes a component defined up to an infinitely reducible symmetry. This hampers the straightforward way to writing the path integral measure, but suggests an alternative based on using the BRST charge and free conformal field theory action of the pure spinor formalism for a step-by-step procedure to find, starting from the solution of pure spinor constraints, the proper set of irreducible variables describing the degrees of freedom of the nonminimal pure spinor formalism.

This paper is organized as follows. In Sec. \ref{SecMin} we briefly (and  schematically) review the  pure spinor approach to quantum superstring. Its nonminimal extension is described in Sec. \ref{SecNonMin}. In Sec. \ref{solution} we present the general solution (\cite{IB07:M0}) of the pure spinor constraints in terms of complex SO(8) null--vectors $\Lambda^{+}_q$ and spinor moving frame variables.

The properties of these spinor moving frame variables are described in Sec.  \ref{sMovingF}.   To make them natural, we begin in sec. \ref{MovingF} by describing moving frame formulation of classical Green--Schwarz superstring and then, in section \ref{sMovingF1}, generalize it till twistor--like or spinor moving frame formulation \cite{BZ-str}. In sec. \ref{sMovingF0} we review the spinor moving frame formulation of the superparticle \cite{IB+AN=95}, where the spinor moving frame variable can be considered as coordinates of the coset of the Lorentz group isomorphic to ${\bb S}^8$, which can be identified with the celestial sphere of a 10D observer (see \cite{Ghsds}). We also discuss there a reformulation of the twistor--like action for superstring in which the generalized spinor moving frame sector parameterize the direct product ${\bb S}^8\times {\bb S}^8$ (instead of the noncompact coset ${Spin(1,9)\over SO(1,1)\times SO(8)}$ in   \cite{BZ-str}).

Not everything in Sec. 5 is needed to understand the rest of the paper; if reader is not interested in technical details, he/she can pass from Sec. 4 directly to Sec. 6, turning to selected equations from sec. 5 if/when necessary.

Sec.  \ref{DuDv=} contains a technical material on derivatives of the spinor moving frame variables which is then used in  Sec. \ref{d11l=sec} to find the expression for the pure spinor measure $d^{11}\lambda$. The first of such  expressions, found in sec. \ref{d11l=subsec}, is further reformulated in sec.  \ref{SO8=CF}  with the use of Cartan forms corresponding to a coset of $SO(8,{\bb C})$ group. That, together with connection-type Cartan forms, provide a basis of cotangent space to the $SO(8,{\bb C})$ group manifold parametrized by complex null--vector $\Lambda^{+}_q$, entering the solution of the pure spinor constraint for $\lambda^\alpha$, its complementary complex null vector $\Lambda^{-}_q$, entering the solution of the constraint for the 'conjugate pure spinor' ${\bar \lambda}_\alpha$, and six complex  vectors, $\Lambda^{I}_q$, orthogonal to those and among themselves. This set of variables helps us to solve, in
sec. \ref{r=sol}, the constraints for the fermionic spinor  $r_\alpha$ of the nonminimal pure spinor formalism. In sec. \ref{d11bL=sec} we present the measure $d^{11}\bar{\lambda}$ for integration over the space of conjugate pure spinors. However, as we discuss in Sec. \ref{d11r=problem}, to write in the similar term the measure $d^{11}r$ for the fermionic pure spinor $r_\alpha$ is not so straightforward because our solution of the constraint for $r_\alpha$ is infinitely reducible; this is to say it is written in terms of larger number of parameters ($15 >11$) restricted by an infinitely  reducible symmetry. However, the previous stages of our procedure suggest a possible way out which we discuss in concluding sec. \ref{discussion}. There we  present the wanted expression for  $d^{11}r$   and discuss a possible reformulation of the pure spinor path integral description of quantum superstring in terms of irreducible variables.

\section{Pure spinor description of superstring I. Minimal pure spinor formalism}\label{SecMin}

In this section we give a very brief and schematic description of {\it minimal} pure spinor approach to quantum superstring \cite{Berkovits:2000fe,Berkovits:2004px}, which is used to describe the tree superstring amplitudes. The main object of this formalism is the BRST charge of the following very simple form
\begin{eqnarray}\label{Q=}
Q= \int \lambda^\alpha D_\alpha \; .   \qquad
\end{eqnarray}
Here $\lambda^\alpha$ is a bosonic spinor and $D_\alpha$ is a quantum fermionic operator representing the fermionic (Grassmann--odd) constraints characteristic of the Green--Schwartz superstring (and, hence, commuting with $\lambda^\alpha$, as this is not a dynamical variable of Green--Schwartz formulation of superstring) and  $\alpha=1,...,16$  is the 10D Weyl spinor index. As 10D Weyl spinor representation has no counterpart of the charge conjugation matrices, the upper spinor index cannot be lowered so that one can state that
$D_\alpha$  carries the left chiral spinor representation of $SO(1,9)$ {\bf 16}$_L$, while the pure spinor $\lambda^\alpha$  carries the right chiral spinor representation {\bf 16}$_R$.

The fermionic constraints $D_\alpha$ anti-commute on a composed bosonic vector field ${\cal P}_a$
\begin{eqnarray}\label{DD=sP}
\{ D_\alpha (\sigma ) \, , \, D_\beta (\sigma^\prime)\}= -2i\sigma^a_{\alpha\beta} {\cal P}_a \delta (\sigma -\sigma^\prime) \; .   \qquad
\end{eqnarray}
The explicit form of ${\cal P}_a $ is again not essential for our discussion here;  $\sigma^a_{\alpha\beta}=\sigma^a_{\beta\alpha}$ is the 10D counterpart of the relativistic Pauli matrices or Klebsh Gordan coefficients for {\bf 10} in the decomposition of the product of two left chiral spinor representations {\bf 16$_L\times$ 16$_L$= 10+120+126}. These obey
\begin{eqnarray}\label{sts=2eta}
(\sigma^a\tilde{\sigma}{}^b+\sigma^b\tilde{\sigma}{}^a)_{\alpha}{}^{\beta}=
2\eta^{ab}\delta _{\alpha}{}^{\beta}\; ,   \qquad
\end{eqnarray} where $\eta^{ab}=diag(+1,-1,..,-1)$ is the 10D  Minkowski space metric and  $\tilde{\sigma}{}_a^{\alpha\beta}=\tilde{\sigma}{}_a^{\beta\alpha}$ is the 'dual' 10D Pauli matrix, i.e. the Klebsh Gordan coefficients for {\bf 10} in the decomposition of the product of two right chiral spinor representations {\bf 16$_R\times$ 16$_R$= 10+120+126}.

The custom of the pure spinor approach literature is to use (instead the Poisson brackets or anti-commutators) the OPE's (operator product expansion) in terms of which (\ref{DD=sP}) is represented by
\begin{eqnarray}\label{DD=OPE}
D_\alpha (z) D_\beta (y)\quad \mapsto \quad {1\over y-z }\sigma^a_{\alpha\beta} {\cal P}_a  \; .   \qquad
\end{eqnarray}

Using this, or Eq. (\ref{DD=sP}), one easily calculates $Q^2= -i \int \lambda \sigma^a \lambda \, {\cal P}_a$ and the key observation is that, if we impose on the bosonic spinor $\lambda^\alpha$ the constraint
\begin{eqnarray}\label{lsl=0}
 \lambda\sigma^a\lambda:= \lambda^\alpha \sigma^a_{\alpha\beta}\lambda^\beta =0 \; ,
\end{eqnarray}
then the BRST charge (\ref{Q=}) is nilpotent
\begin{eqnarray}\label{Q^2=0}
 Q^2=0 \qquad \Leftarrow \qquad \lambda\sigma^a\lambda=0 \; .
\end{eqnarray}

Spinor obeying the constraint (\ref{lsl=0}) is called 'pure spinor'. The notion of pure spinor was introduced by Cartan \cite{Cartan}. In the context of supersymmetric theories the D=10 pure spinors were introduced and used in \cite{Nilsson:1985cm,Howe:1991mf,Howe:1991bx}. The treatment of the nilpotent operator (\ref{Q=}) as BRST charge suggests that in the pure spinor approach to superstring the pure spinor is considered as a ghost variable. This also advocates to require the wavefunctions and vertex operator to depend on $\lambda$ polynomially.

Notice that the pure spinors defined in such a way are complex (this is to say: Weyl but not Majorana--Weyl), as far as for real (Mayorana--Weyl) 10D spinors  the constraint (\ref{lsl=0}) has only trivial solution $\lambda_{MW}=0$. For complex (Weyl) spinors the space of solutions of the $D=10$ pure spinor constraint (\ref{lsl=0}) is 11-dimensional \cite{Berkovits:2000fe}.

Another ingredient of the minimal pure spinor formalism is the free conformal field theory (CFT) action for all (the left moving\footnote{After Wick rotation, usually assumed in the pure spinor approach, the left-moving fields become holomorphic and right-moving become anti--holomorphic.}) degrees of freedom
\begin{eqnarray}\label{Smin=}
S_{min}= \int \left(1/2 \partial x^\mu \bar{\partial} x^\mu + p_\alpha \bar{\partial} \theta^\alpha - w_\alpha \bar{\partial} \lambda^\alpha \right)
\; . \qquad
\end{eqnarray}
Here $p_\alpha $ is momentum for the (left-moving) Grassmann coordinate function $\theta^\alpha $, a superpartner of the (left moving part of the) bosonic coordinate function $x^\mu$ (this enters (\ref{Smin=}) together with its right-moving counterpart) and $w_\alpha$ is the momentum conjugate to the pure spinor $\lambda^\alpha$. The fact that this later is constrained by (\ref{lsl=0}) is reflected by the gauge symmetry acting on its momentum,
\begin{eqnarray}\label{vw=}
\delta w_\alpha = \Xi^a (\sigma_a\lambda)_\alpha
\; . \qquad
\end{eqnarray}
This symmetry leaves invariant the set of currents ($\sigma_{ab\alpha}{}^\beta:= {1\over 2}(\sigma_{a\alpha\gamma}\tilde{\sigma}_{b}^{\gamma\beta}-\sigma_{b\alpha\gamma}\tilde{\sigma}_{a}^{\gamma\beta} )$)
\begin{eqnarray}\label{N=}
N_{ab}:={1\over 2}\lambda \sigma_{ab}w\; , \qquad J_{\lambda}:= \lambda w \; , \qquad \\ \label{T=}
T_{\lambda}:= \partial\lambda w
\; , \qquad
\end{eqnarray}
the last of which, (\ref{T=}), is the 2d energy momentum tensor of the pure spinor field.

To avoid the doubling of the degrees of freedom, one imposes a requirement of analyticity, this is to say, to consider all the  wavefunctions and amplitudes to have holomorphic dependence on $\lambda^\alpha$ ({\it i.e.}, to be dependent  on $\lambda^\alpha$  but not on $(\lambda^\alpha)^*$). Then the superstring path integral \cite{Berkovits:2004px,Berkovits:2005bt,Berkovits:2006vi} contains  a chiral measure, which is based on the 11--form $d^{11}\lambda$ (rather than on 22-form $d^{11}\lambda \wedge (d^{11}{\lambda})^*$).

A (relatively) simple expression for this 11 form  \cite{Mafra:2009wi,Mafra:2009wq} (which is implicit in \cite{Berkovits:2005bt,Berkovits:2006vi})
\begin{eqnarray}\label{d11L=Berk}
d^{11}\lambda = {1\over (\bar{\lambda}\lambda)^{^3}}\; (\bar{\lambda}\tilde{\sigma}{}^{\underline{a}})^{\alpha_1}(\bar{\lambda}\tilde{\sigma}{}^{\underline{b}})^{\alpha_2}(\bar{\lambda}\tilde{\sigma}{}^{\underline{c}})^{\alpha_3}
(\tilde{\sigma}_{\underline{a}\underline{b}\underline{c}})^{\alpha_4\alpha_5}\epsilon_{\alpha_1 \ldots \alpha_5\beta_1 \ldots \beta_{11}} d\lambda^{\beta_1}\wedge \ldots \wedge d\lambda^{\beta_{11}}\; , \qquad
\end{eqnarray}
is written with the use of a dual or ''conjugate'' spinor $\bar{\lambda}_\alpha$ restricted by the condition that
\begin{eqnarray}\label{not=0}
(\lambda \bar{\lambda}) \equiv \lambda^\alpha \bar{\lambda}_\alpha \not=0
\; . \qquad
\end{eqnarray}
This can be considered just as a reference spinor in the minimal pure spinor formalism, but appeared to be a necessary ingredient in the non--minimal pure spinor formalism introduced in \cite{Berkovits:2005bt}.

\section{Pure spinor description of superstring II. Nonminimal pure spinor formalism}\label{SecNonMin}

The non--minimal formalism \cite{Berkovits:2005bt} includes, in addition to the coordinate functions and the pure spinor $\lambda^\alpha$, also the above dual or conjugate spinor $\bar{\lambda}_\alpha$ obeying the pure spinor constraint
\begin{eqnarray}\label{bltsbl=0}
 \bar{\lambda}\tilde{\sigma}{}_a\bar{\lambda}:= \bar{\lambda}_\alpha \tilde{\sigma}{}_a^{\alpha\beta}\bar{\lambda}_\beta =0 \; ,
\end{eqnarray}
as well as the fermionic spinor $r_\alpha$, obeying
\begin{eqnarray}\label{bltsr=0}
 \bar{\lambda}\tilde{\sigma}_ar:= \bar{\lambda}_\alpha \tilde{\sigma}{}_a^{\alpha\beta}r_\beta =0 \; .
\end{eqnarray}
It also uses their canonically conjugate momenta: bosonic $\bar{w}^\alpha$ and fermionic $s^\alpha$. Due to the constraints
(\ref{bltsbl=0}) and (\ref{bltsr=0}) these are defined up to the gauge transformations
\begin{eqnarray}\label{vbw=}
\delta \bar{w}^\alpha = \bar{\Xi}{}^a (\tilde{\sigma}_a\bar{\lambda})^\alpha - {\xi}{}^a (\tilde{\sigma}_ar)^\alpha
\; , \qquad \delta s^\alpha = {\xi}{}^a (\tilde{\sigma}_a\bar{\lambda})^\alpha
\; . \qquad
\end{eqnarray}
These leave invariant the (free CFT) action of the nonminimal pure spinor formalism,
\begin{eqnarray}\label{Snmin=}
S_{nonmin}= \int \left(1/2 \partial x^\mu \bar{\partial} x^\mu + p_\alpha \bar{\partial} \theta^\alpha - w_\alpha \bar{\partial} \lambda^\alpha -  \bar{w}^\alpha \bar{\partial} \bar{\lambda}_\alpha + s^\alpha \bar{\partial} r_\alpha \right)
\; . \qquad
\end{eqnarray}
and also the BRST charge of the nonminimal pure spinor formalism,
\begin{eqnarray}\label{Q=lD+rw}
Q= \int (\lambda^\alpha D_\alpha  + \bar{w}^\alpha r_\alpha)\; ,   \qquad
\end{eqnarray}
which contains, besides (\ref{Q=lD+rw}),  the additional contribution of the nonminimal sector, $\int \bar{w}^\alpha r_\alpha$.

The new momentum variables, bosonic $\bar{w}^\alpha$ and fermionic $s^\alpha$, appear inside the five combinations which include new  contributions to the bosonic currents  (\ref{N=}) and (\ref{T=}),
\begin{eqnarray}\label{N=bl}
\bar{N}_{ab}:={1\over 2}\left(\bar{w} \sigma_{ab}\bar{\lambda} - s\sigma_{ab}r\right)\; , \qquad \bar{J}_{\bar{\lambda}}:= \bar{w}\bar{\lambda}-sr  \; , \qquad
\\ \label{T=bl}
T_{\bar{\lambda}}:= \bar{w}\partial\bar{\lambda} -s\partial r
\; , \qquad
\end{eqnarray}
and the fermionic currents
\begin{eqnarray}\label{S=bl}
{S}_{ab}:={1\over 2} s \sigma_{ab} \bar{\lambda} \; , \qquad S:=  s  \bar{\lambda}
\; . \qquad
\end{eqnarray}

The path integral of the nonminimal pure spinor formalism includes the integration over  the fields  $\bar{\lambda}_\alpha$ and  $r_\alpha$ which are based on the  the measures \cite{Mafra:2009wi}
\begin{eqnarray}\label{d11bL=}
d^{11}\bar{\lambda} = {1\over (\bar{\lambda}\lambda)^{^3}}\; ({\lambda}{\sigma}{}^{\underline{a}})_{\alpha_1}({\lambda}{\sigma}{}^{\underline{b}})_{\alpha_2}
({\lambda}{\sigma}{}^{\underline{c}})_{\alpha_3}
\, {\sigma}_{\underline{a}\underline{b}\underline{c}\; \alpha_4\alpha_5}\epsilon^{\alpha_1 \ldots \alpha_5\beta_1 \ldots \beta_{11}} d\bar{\lambda}_{\beta_1}\wedge \ldots \wedge d\bar{\lambda}_{\beta_{11}}
\;    \qquad
\end{eqnarray}
and
\begin{eqnarray}\label{d11r=}
d^{11}r=
{1\over (\bar{\lambda}\lambda)^{^3}}\; (\bar{\lambda}\tilde{\sigma}{}^{\underline{a}})^{\alpha_1}(\bar{\lambda}\tilde{\sigma}{}^{\underline{b}})^{\alpha_2}(\bar{\lambda}\tilde{\sigma}{}^{\underline{c}})^{\alpha_3}
(\tilde{\sigma}_{\underline{a}\underline{b}\underline{c}})^{\alpha_4\alpha_5}\epsilon_{\alpha_1 \ldots \alpha_5\beta_1 \ldots \beta_{11}} {\partial\over \partial r_{\beta_1}}\ldots {\partial\over \partial r_{\beta_{11}}}\; . \qquad
\end{eqnarray}

To understand better the structure of the pure spinor path integral for quantum superstring, it may be useful to find another, although equivalent, form of the measure factors (\ref{d11L=Berk}), (\ref{d11bL=}), (\ref{d11r=}). In the recent \cite{Grassi:2011ie} it was studied the possibility to extract the pure spinor measure $d^{11}\lambda$ from the ${\mathbf C}^{16}-\{0\}$ integration measure of the space of nonvanishing unconstrained 10D Weyl  spinors. Here we will try to go in an opposite way, trying to express the measure, and also  the fields of the pure spinor formalism, in terms of constrained but irreducible variables; this is to say we will try to solve the pure spinor condition and other constraints of the pure spinor formalism in such a way that no reducible symmetries appear.

\section{Pure spinors and spinor moving frame} \label{solution}

We begin by the general solution of the $D=10$ pure spinor constraint (\ref{lsl=0})
which was found in \cite{IB07:M0},
\begin{eqnarray}\label{l=L+v-}
\lambda^\alpha =\Lambda^{+}_q v_q^{-\alpha} \; , \qquad \Lambda^{+}_q\Lambda^{+}_q=0 \; . \qquad
\end{eqnarray}
This involves a  complex null vector
$\Lambda^{+}_q$ (the measure in the space of which was considered in the appendix of \cite{Grassi:2011ie}), which carries the ghost number characterizing the pure spinor $\lambda^\alpha$, and also the set of 8 Majorana--Weyl spinors constrained by
\begin{eqnarray}\label{v-tsv-=u--}
& v_{{q}}^{-}{\sigma}^{{a}}v_{{p}}^{-}=
\delta_{{q}{p}} u^{=}_{{a}}& \; , \qquad
u^{=}_{{a}}u^{={a}}=0 \; , \qquad \\ \label{v-qs5v-q=0}
&v_{{q}}^{-}{\sigma}^{{a}_1...a_5}v_{{q}}^{-}=0 &\; .
\end{eqnarray}
Due to the first equation in (\ref{v-tsv-=u--}),  the composite  spinor  (\ref{l=L+v-}) obeys $\lambda\sigma^a\lambda = \Lambda^+_q\Lambda^+_q u^{=a}$ and the {\it r.h.s.} of this equation vanishes due to that the complex 8 component $\Lambda^+_q$ is null vector,  $\Lambda^+_q\Lambda^+_q=0$. Thus (\ref{l=L+v-}) indeed solves (\ref{lsl=0}); moreover, it is the general solution of (\ref{lsl=0}).

Indeed, as we discuss  below in more detail (see sec. \ref{d11l=subsec}), the variables in the {\it r.h.s.} of Eq.(\ref{l=L+v-}) carry 11 complex or 22 real not-pure-gauge degrees of freedom (d.o.f.-s), the same as the number of d.o.f.-s in the 10D pure spinor $\lambda^\alpha$ \cite{Berkovits:2000fe}. Of these the 7 complex d.o.f.-s are carried by the complex null 8-vector $\Lambda^+_q$ and 8 real (4 complex) are the not-pure gauge d.o.f.-s in $v_q^{-\alpha}$.

In short, to understand this it is important to notice first that, up to the scaling transformations
\begin{eqnarray}\label{GL1}
u^{={a}}\mapsto u^{={a}}e^{-2\beta}
\; , \qquad
\end{eqnarray}
the light--like 10-vector $u_a^{=}$ (in (\ref{v-tsv-=u--})) parametrizes eight--sphere \footnote{One can solve the ligh--likeness constraints by  $u_a^{=}= E (1, n^{{I}})$ where $I=1,...,9$ and $n^{{I}}$ satisfy $n^In^I=1$ and hence parametrizes ${\bb S}^8$. When $u_a^{=}$ is related to the momentum of massless particle, this ${\bb S}^8$ can be recognized as celestial  sphere.
} ${\bb S}^8$. Then (and this is
what  makes the solution (\ref{l=L+v-}) of the pure constraint interesting) one can show that  the space  of 8 spinors $v_q^{-\alpha}$ constrained by (\ref{v-tsv-=u--}), (\ref{v-qs5v-q=0}) {\it and} considered modulo $SO(8)$ and scaling transformations,
\begin{eqnarray}\label{GL1}
v_q^{-\alpha}\mapsto {\cal O}_{pq}v_p^{-\alpha} e^{-2\beta}
\; , \qquad  {\cal O}_{pq'} {\cal O}_{q'q}=\delta_{pq}\; , \qquad
\end{eqnarray}
is an  eight--dimensional space which doubly covers the eight sphere ${\bb S}^8$.

The easiest way to see this implies  considering  the light--like vector $u_a^{=}$ as an element of a {\it moving frame} and the set of constrained spinors $v_q^{-\alpha}$ as an element of {\it spinor moving frame}. Although to this end we introduce a new set of constrained variables, this is convenient  because in such a moving frame formulation it is possible to find a covariant set of irreducible constraints defining the basic variables  \cite{BZ-smem} and thus to escape the necessity to use the reducible constraints and reducible symmetries.

\section{Moving frame, spinor moving frame and a twistor--like formulation  of classical superstring}\label{sMovingF}

This section contains a brief review on moving frame and spinor moving frame variables and their applications to classical description of superstring.

Not everything described in this section is strictly necessary to understand the rest of the paper; the reader who is not interested in technical  details of spinor moving frame formalism  can pass directly to Sec. 6, turning to selected equations of this section if/when necessary.

The moving frame associated to a massless (super)particle and (super)string is described by the $SO(1,9)$ valued matrix
\begin{eqnarray}\label{harmUin}
& U_{{a}}^{(b)}= \left( {1\over 2}\left(u_a^{\#}+u_a^{=}\right), \; u_a^{j}, \; {1\over 2}\left(u_a^{\#}+u_a^{=}\right)\right)\;  \in \;
SO(1,9)\; .  \qquad
\end{eqnarray}
The condition (\ref{harmUin}) implies $U^T\eta U=\eta=diag(+1,-1,...,-1)$ or, in more detail,
 \begin{eqnarray}\label{harmU10}   \left\{\matrix{
u_{{a}}^{=}u^{{{a}}=}=0 \; ,  \quad
& u_{{a}}^{\#}u^{{{a}}\#}=0 \; , \qquad
& u_{{a}}^{=}u^{{{a}}\#}=2 \; , \quad  \cr
& u_{{a}}^{=}u^{{{a}}\, i}=0\; , \quad & u_{{a}}^{\#}u^{{{a}}\, i} =0\; , \quad \cr
&&  u_{{a}}^{i}u^{{{a}}\, j}=- \delta^{ij}\; , \quad }\right. \qquad
\end{eqnarray}
as well as $U\eta U^T=\eta$ which reads as a unity decomposition
\begin{eqnarray}\label{Udec}
\delta_{\underline{a}}{}^{\underline{b}}= {1\over
2}u_{\underline{a}}^{\#}u^{\underline{b}=} + {1\over
2}u_{\underline{a}}^{=}u^{{\underline{b}}\#} -
u_{\underline{a}}^{i}u^{{\underline{b}} i}\; .
\end{eqnarray}
The decomposition in  (\ref{harmUin})  is invariant under  the left action of the
10D Lorentz group $SO(1,9)$ and under the right action of $SO(1,1)\times SO(8)$ subgroup of $SO(1,9)$. If dynamical model under study possesses gauge symmetry under these right $SO(1,1)\times SO(8)$ transformations, one can use them as an identification relation in the SO(1,9) group manifold parametrized by the vectors (\ref{harmUin}) constrained by (\ref{harmU10}). Then one can consider these constrained vectors as homogenous coordinates of the  $SO(1,9)/[SO(1,1)\times SO(8)]$ coset (see \cite{BZ-str}).
This makes  moving frame variables similar to the so--called harmonic variables of the R-symmetry groups \cite{GIKOS} useful to formulate the $N=2,3$ supersymmetric theory in terms of unconstrained superfields. Such a similarity was the reason  to call the moving frame and spinor moving frame variable 'Lorentz harmonics' ('spinorial harmonics') \cite{B90,BZ-str}; also the name 'light-cone harmonics' was used in \cite{Sok}.

\subsection{Moving frame formulation of superstring}\label{MovingF}

The Goldstone fields in the coset $SO(1,9)/[SO(1,1)\times SO(8)]$ describe the spontaneous breaking of $SO(1,9)$ symmetry down to its $[SO(1,1)\times SO(8)]$  subgroup. Such a symmetry breaking is characteristic for a 10D string (and superstring) model so that it is not surprising that one can write a formulation of superstring action with the use of the moving frame variables. Such an action was proposed and studied in \cite{BZ-str}. In an arbitrary 10D supergravity background the action of the moving frame formulation can be written as
\begin{equation}\label{SIIB-2ord}
S_{moving\; frame}=   {1 \over 2} \int_{W^{2}}  \hat{E}^{\#} \wedge
\hat{E}^{=} - \int_{W^{2}} \hat{B}_2 \; ,
\end{equation}
where $\wedge$ is the exterior product of differential forms ($\hat{E}^{\#} \wedge
\hat{E}^{=} =-
\hat{E}^{=} \wedge \hat{E}^{\#} $),
\begin{eqnarray}\label{E++E--=}
\hat{E}^{\#}:= \hat{E}^{{a}}u_{{a}}^{\#}\; ,
\qquad \hat{E}^{=}:=
\hat{E}^{{a}}u_{{a}}^{=}\; , \qquad
\end{eqnarray}
where $u_{{a}}^{\#}(\tau, \sigma) $ and $u_{{a}}^{=}(\tau, \sigma)$ are light--like moving frame vector fields obeying (\ref{harmU10}) and
\begin{eqnarray}\label{hEa=}
\hat{E}^{{a}}:= d\hat{Z}^M(\xi) {E}_M{}^{{a}} (\hat{Z}(\xi))=d\xi^m \hat{E}_m^{{a}}\; , \qquad
\hat{E}_m^{{a}} =
 \partial_m \hat{Z}^M(\xi)
{E}_M{}^{{a}} (\hat{Z}(\xi))  \qquad
\end{eqnarray} is the pull--back of the bosonic supervielbein of the 10D  superspace,
$E^a(Z):=dZ^ME_M{}^a(Z)$, to the worldsheet $W^2$ with local coordinates $\xi^m=(\tau, \sigma)$. In the case of flat $N=1$ superspace
\begin{eqnarray}\label{hEa=}
E^a= \Pi^\mu\delta_\mu^a= dx^\mu \delta_\mu^a +{1\over 2}\theta^\alpha \sigma^a_{\alpha\beta}d\theta^\beta
 \qquad
\end{eqnarray}
(here we try to follow the notation of pure spinor literature \cite{Berkovits:2000fe,Berkovits:2004px,Berkovits:2005bt,Berkovits:2006vi,Berkovits:2006ik,Mafra:2009wq}). Finally,
$-\int_{{\cal M}^{2}} \hat{B}_2 $ is the Wess--Zumino term, the same as in the original
Green--Schwarz formulation of the superstring action; that reads
\begin{eqnarray}\label{SGS=}
S_{GS}=   {1 \over 2} \int_{W^{2}} d^2\xi \sqrt{|det(g_{mn})|}  - \int_{W^{2}} \hat{B}_2\qquad
\end{eqnarray}
 with $g_{mn}:= \hat{E}_m^a\eta_{ab}\hat{E}_n^b$ and $d^2\xi=d\tau \wedge d\sigma$.
We do not need in explicit form of $B_2={1\over 2}dZ^M\wedge dZ^NB_{NM}(Z)$ in this paper.

One can also introduce auxiliary worldsheet vielbein  $e^{\#}=d\xi^m e^{\#}_m(\xi)$, $e^{=}=d\xi^m e^{=}_m(\xi)$
and write the moving frame action in the first order form  \cite{BZ-str},
\footnote{The formulations of superstring model proposed and developed in \cite{niss}
and \cite{K+R88} introduced Lorentz harmonics (moving frame variables and (counterparts of)  spinor moving frame variables)
as additional variables in the Hamiltonian approach to superstring developed on the basis of the
standard Lagrangian action (\ref{SGS=}).
}
\begin{eqnarray}\label{SIIB-1ord}
& S^\prime_{moving\; frame}= {1 \over 2} \int_{W^{2}} \left( {e}^{\#}
\wedge \hat{E}^{=} - {e}^{=} \wedge \hat{E}^{\#}
 - {e}^{\#} \wedge
{e}^{=} \right) - \int_{W^{2}} \hat{B}_2 \; .
\end{eqnarray}

\subsection{Spinor moving frame}\label{sMovingF1}

The above moving frame formulation of superstring, characterized by the action (\ref{SIIB-1ord}) and (\ref{SIIB-1ord}),  was also called 'twistor--like' and  'spinor moving frame' formulation. This is related to the fact that one can introduce a set of constrained spinors $v_q^{-\alpha }$ which
can be considered as square roots of the light--like vector
$u_a^{=}$ in the sense of (the trace part of) Eq. (\ref{v-tsv-=u--});
similarly, one can introduce a set of constrained spinors $v_{\dot{q}}^{+\alpha}$
 which can be considered as square roots of the light--like vector $u_a^{\#}$,
\begin{eqnarray}\label{v+tsv+=u++}
&& v_{\dot{q}}^{+}{\sigma}^{{a}}v_{\dot{p}}^{+}=
\delta_{\dot{q}\dot{p}} \, u^{\#}_{{a}} \; , \qquad
u^{\#}_{{a}}u^{\#{a}}=0 \; , \qquad \\ \label{v+dqs5v+dq=0}
&& v_{\dot{q}}^{+}{\sigma}^{abcde}v_{\dot{q}}^{+}=0\; . \qquad
\end{eqnarray}

These two sets of constrained spinors can be considered as $16\times 8$ blocks
of the spinor moving frame matrix
\begin{eqnarray}\label{V-1STR}
 V_{(\alpha)}{}^{\beta} = (v^{- {\beta}}_q\; , v^{+ {\beta}}_{\dot{q}} ) \; \in \;
Spin(1,9)\; .
\end{eqnarray}
The $Spin(1,9)$ valuedness can be expressed as a statement of  Lorentz invariance of the 10D Pauli matrices
$\sigma^a_{\alpha\beta}=\sigma^a_{\beta\alpha}$  and $\tilde{\sigma}{}^{a\,\alpha\beta}=
\tilde{\sigma}{}^{a\,\beta\alpha}$  which obey
\begin{eqnarray}\label{sts+sts=I}
\sigma^a\tilde{\sigma}{}^b+ \sigma^b\tilde{\sigma}{}^a=\eta^{ab}I_{16\times 16}
\; .
\end{eqnarray}
Namely, Eq. (\ref{V-1STR}) ($V\in Spin(1,9)$) states that the similarity transformation of the 10D
Pauli matrices with the matrix $V$ results in a linear combination of these Pauli matrices,
\begin{eqnarray}\label{VsV=sU}
V{\sigma}_{{a}} V^T =
U_{{a}}^{({b})} {\sigma}_{({b})}\; , \qquad
  V^T \tilde{\sigma}^{({a})} V =
U_{{b}}^{({a})} \tilde{\sigma}^{{b}}\; .
 \qquad
\end{eqnarray}
The $10\times 10$ matrix of the coefficients in the {\it r.h.s.} of Eqs. (\ref{VsV=sU}) takes its values in the Lorentz group $SO(1,9)$. Identifying it with the moving frame matrix (\ref{harmU10}) we find that (\ref{VsV=sU}) implies (\ref{v-tsv-=u--}), (\ref{v+tsv+=u++}) and a set of similar relations,
\begin{eqnarray}\label{v+sv+=u++}
v_{\dot{q}}^{+}{\sigma}^{{a}}v_{\dot{p}}^{+}=
\delta_{\dot{q}\dot{p}} u^{\#}_{\underline{a} } \; , & \qquad &
 u^{\#}_{\underline{a}}
\tilde{\sigma}^{\underline{a}{\alpha}{\beta}} = 2v^{+{\alpha}}_{\dot{q}}
v_{\dot{q}}^{+{\beta}}
\; , \qquad \\
\label{v-sv-=u--}
v_{{q}}^{-}{\sigma}^{\underline{a}}v_{{p}}^{-}=
\delta_{{q}{p}} u^{=}_{\underline{a}} \; ,  & \qquad &
 u^{=}_{\underline{a}}
\tilde{\sigma}^{\underline{a}{\gamma}{\beta}} = 2v^{-{\gamma}}_{q} v_{q}^{-{\beta}}\; ,
\qquad \\ \label{v-sv+=ui}
v_{{q}}^{-}{\sigma}^{\underline{a}}v_{\dot{q}}^{+}= -
\gamma^i_{{q}\dot{q}} u^{i}_{\underline{a}} & \qquad &
 u^{i}_{\underline{a}}
\tilde{\sigma}^{\underline{a}{\gamma}{\beta}} = - v^{-{\gamma}}_{{q}}
\gamma^i_{q\dot{q}} v_{\dot{q}}^{+{\beta}} - v^{-{\beta}}_{{q}} \gamma^i_{q\dot{q}}
v_{\dot{q}}^{+{\gamma}}\; , \qquad \; . \qquad
\end{eqnarray}
Notice that (\ref{v-qs5v-q=0}) and (\ref{v+dqs5v+dq=0}) are obeyed as consequences of the second equation in (\ref{v+sv+=u++}) and of the second equation in (\ref{v-sv-=u--}), respectively.

To write the similar square root type relation in terms of matrices $u^{=}_{\underline{a}}
\sigma^{\underline{a}}_{{\alpha}{\beta}}$ and $u^{\#}_{\underline{a}}
\sigma^{\underline{a}}_{{\alpha}{\beta}}$, one has to introduce the inverse spinor moving frame matrix
\begin{eqnarray}\label{VSTRinG-H}
 V_\alpha^{(\beta)} = \{ (v_{\alpha q}^{\;\,+}\; , v_{\alpha\dot
q}^{\;\,-})\} \quad \in Spin(1,9) \; ,
\qquad
\label{V-1:=}
 V_{(\alpha)}^{\;\;\;
\gamma} V_{\gamma}^{(\beta)} =\delta_{(\alpha)}^{\;\; (\beta)}:= \left(\matrix{
\delta_{{q}}^{{p}} & 0\cr 0 & \delta_{\dot{q}}^{\dot{p}} }\right)\;
\end{eqnarray}
Its elements obey the constraints
\begin{eqnarray}\label{V-1V=ISTR}
v^{-{\alpha}}_{p} v^{~+}_{{\alpha}q}= \delta_{pq}, \qquad v^{-{\alpha}}_{p}
v^{~-}_{{\alpha}\dot{q}} =0\, , \qquad \nonumber
\\
v^{+{\alpha}}_{\dot{p}} v^{~+}_{{\alpha}q}=0, \qquad v^{+{\alpha}}_{\dot{p}}
v^{~-}_{{\alpha}\dot{q}} =\delta_{\dot{p}\dot{q}}\; \qquad
\end{eqnarray}
and also
\begin{eqnarray}\label{u++=v+v+2}
 u^{\#}_{\underline{a}}
\sigma^{\underline{a}}_{{\alpha}{\beta}} = 2v_{{\alpha}q}^{~+} v^{~+}_{{\beta}q}\; , &
\qquad & v_{q}^{+}\tilde{\sigma}^{\underline{a}}v_{p}^{+}= \delta_{qp}
u^{\#}_{\underline{a} }\; , \qquad
\\ \label{u--=v-v-2}
 u^{=}_{\underline{a}}
\sigma^{\underline{a}}_{{\alpha}{\beta}} = 2v_{{\alpha}\dot{q}}^{~-}
v^{~-}_{{\beta}\dot{q}}\; , &
\qquad & v_{\dot{q}}^{-}\tilde{\sigma}^{\underline{a}}v_{\dot{p}}^{-}=
\delta_{qp} u^{=}_{\underline{a}} \; ,
\qquad
\\ \label{ui=v+v-2}
 u^{i}_{\underline{a}}
\sigma^{\underline{a}}_{{\alpha}{\beta}} = 2v_{({\alpha}{q}}^{~+} \gamma^i_{q\dot{q}}
v^{~-}_{{\beta})\dot{q}}\; , & \qquad &
v_{{q}}^{-}{\sigma}^{\underline{a}}v_{\dot{q}}^{+}= -
\gamma^i_{{q}\dot{q}} u^{i}_{\underline{a}} \; . \qquad
\end{eqnarray}

\subsection{Spinor moving frame and celestial sphere}\label{sMovingF0}

The (spinor) moving frame formulation of the massless superparticle
is based on the action \cite{B90,IB+AN=95}
\begin{eqnarray}\label{Smassless=}
& S^\prime_{}=  \int_{W^{1}} {\rho}^{\#} \hat{E}^{=}= \int d\tau {\rho}^{\#}(\tau)
\hat{E}^{a}u_a^{=}(\tau)  \;
\end{eqnarray}
involving only one light--like spinor moving frame vector $u_a^{=}$. Then the
natural SO(1,1)$\otimes$SO(8) gauge symmetry of the splitting (\ref{harmUin})
in the superparticle action is enlarged till the semi-direct product  $[SO(1,1)\otimes SO(8)] \subset\!\!\!\!\!\!\times{K}_8$ \cite{Ghsds}, where  $K_8$ transformations act on the moving frame as
\begin{eqnarray}
\label{vK8u=}
 && \delta u^{=}_a=0\; , \qquad \delta u^{\# }_a= 2k^{\# i} u^{i}_a\; , \qquad \delta u^{i}_a=
k^{\# i} u^{=}_a\; , \qquad
\end{eqnarray}
and on the spinor moving frame variables by
\begin{eqnarray}
\label{vK8v=}
&& \delta v_q^{-\alpha} =0 \; , \qquad \delta v_{\dot{q}}^{+\alpha} = - k^{\# i}
v_p^{-\alpha} \gamma_{p\dot{q}}^{i}
\; . \qquad
\end{eqnarray}
As it was stressed in \cite{Ghsds},  $[SO(1,1)\otimes SO(8)\subset\!\!\!\!\!\!\times K_8]$ is the maximal parabolic subgroup of $SO(1,9)$ so that the coset
$Spin(1,9)/[SO(1,1)\otimes SO(8)\subset\!\!\!\!\!\!\times K_8]$ is a compact space which
is actually isomorphic to the 8 sphere ${\bb S}^8$,
\begin{eqnarray}\label{S8}
& {Spin(1,9)\over SO(1,1)\otimes SO(8)\subset\!\!\!\!\times K_8}= {\bb S}^8
\; . \qquad
\end{eqnarray}
In the superparticle model (\ref{Smassless=}) this ${\bb S}^8$ can be identified
with celestial sphere of the  10D observer \cite{Ghsds}.

Furthermore, taking into account Eq. (\ref{u--=v-v-2}) one can write the functional (\ref{Smassless=}) in the form
\begin{eqnarray}\label{D10FSaction=}
& S^\prime_{}=  {1\over 8} \int d\tau {\rho}^{\#}(\tau)
v^-_q\sigma_av^-_q \hat{E}^{a} \;  \qquad
\end{eqnarray}
which makes transparent that we are dealing with 10D generalization of the  D=4 Ferber--Schirafuji action functional \cite{F77S83} which provides a Lagrangian basis for the Penrose twistor approach \cite{Penrose}. This is why the spinor moving frame formulation of superparticle \cite{B90,IB+AN=95} is also called twistor--like.

Using Eq. (\ref{u--=v-v-2}) and (\ref{u++=v+v+2}) one can also write the action  (\ref{SIIB-1ord}) as the sum of two terms similar to (\ref{D10FSaction=}) and a cosmological term; this observation suggested to call 'twistor--like' also the superstring spinor moving frame action  (\ref{SIIB-1ord}) \cite{BZ-str}.

Notice that the superstring model can be formulated not only with moving frame variables
parametrizing the noncompact coset, as in (\ref{SIIB-1ord}),
but also with the action functional
\begin{eqnarray}\label{SIIB-utu}
& S^\prime_{moving\; frame}= {1 \over 2} \int_{W^{2}} \left( {e}^{\#}
\wedge \hat{E}^au_a^{=} - {e}^{=} \wedge \hat{E}^a \tilde{u}_a^{\#}
 - {e}^{\#} \wedge
{e}^{=} \tilde{u}^{a\#}u_a^{=}\right) - \int_{W^{2}} \hat{B}_2 \;
\end{eqnarray}
involving two light--like vectors form different moving frames;
this is to say we do not require the contraction $u_a^{=}\tilde{u}^{a\#}$ to be equal to a fixed constant.
The auxiliary field sector of this model is the direct product of compact spaces,  ${\bb S}^8\otimes {\bb S}^8$.

\subsection{Derivatives of the moving frame variables}\label{DuDv=}

Although the moving frame and spinor moving frame variables are highly constrained,
their transparent  group theoretical meaning allows to calculate easily their derivatives and variations \cite{BZ-str,IB+AN=95}\footnote{The authors of the relatively recent
\cite{GomisWest06} prefer to work instead with explicit solutions of the
constraints on the moving frame variables.}.

Referring to the original references for details, here we just present just results.
The $SO(1,9)\times SO(1,1)\otimes SO(8)$ covariant derivatives of the moving frame and spinor moving frame variables read
\begin{eqnarray}
\label{Du--=}   Du^{=}_a:= & du^{=}_a+ 2u^{=}_a
\Omega^{(0)}  & =   u^{i}_a
\Omega^{=i} \;  , \qquad
 \\
\label{Du++=}   Du^{\#}_a:= & du^{\#}_a- 2u^{\#}_a
\Omega^{(0)} & =  u^{i}_a
\Omega^{\#i} \; , \qquad \\
\label{Dui=}
  Du_a{}^{i} := & du^{i}_a+ u^{j}_a
\Omega^{ji}  & =  {1\over 2}u^{=}_a \Omega^{\# i} + {1\over 2}u^{\#}_a \Omega^{= i} \; .  \qquad
 \end{eqnarray}
Here $\Omega^{\# i}:= u^{\# a}du_a^{i}$ and $\Omega^{= i}:= u^{= a}du_a^{i}$ are covariant Cartan forms providing the vielbein for the coset $SO(1,9)/[SO(1,1)\times SO(8)]$. The Cartan forms  $\Omega^{ij}:= u^{i a}du_a^{j}$ and
 $\Omega^{(0)}:= {1\over 4}u^{= }du_a^{\#}$ transform as $SO(8)$ and $SO(1,1)$ connections.

The $SO(1,1)\otimes SO(8)$ covariant derivatives
of the spinor moving frame  variables is expressed through the same Cartan forms  by
\begin{eqnarray}
\label{Dv-1-q=} &  Dv_q^{-\alpha} := dv_q^{-\alpha} +
\Omega^{(0)} v_q^{-\alpha} + {1\over 4} \Omega^{ij}
v_p^{-\alpha}\gamma_{pq}^{ij} = - {1\over 2} \Omega^{=i}
v_{\dot{p}}^{+\alpha} \gamma_{q\dot{p}}^{i}\; , \qquad \\
\label{Dv-1+dq=} &  Dv_{\dot{q}}^{+\alpha} := dv_{\dot{q}}^{+\alpha} -
\Omega^{(0)} v_{\dot{q}}^{+\alpha} + {1\over 4} \Omega^{ij}
v_{\dot{p}}^{+\alpha}\gamma_{\dot{p}\dot{q}}^{ij} = - {1\over 2} \Omega^{\# i}
v_p^{-\alpha} \gamma_{p\dot{q}}^{i}\;  \qquad
\end{eqnarray}
Now, using (\ref{V-1V=ISTR}), is not hard to find that for the elements of the inverse spinor moving frame matrix the covariant derivative read
\begin{eqnarray}\label{Dv+q=STR}
 Dv_{\alpha {q}}^{\;\;  +}= {1\over 2} \Omega^{\#i}
\gamma_{q\dot{p}}^{i} v_{\alpha \dot{p}}^{\;\; +} \; , \qquad
\label{Dv-dq=STR}  Dv_{\alpha \dot{q}}^{\;\;  -}= {1\over 2}
\Omega^{=i} \; v_{\alpha {p}}^{+}\gamma_{p\dot{q}}^{i}\; .  \qquad
\end{eqnarray}

The differential (exterior derivative) in the space of moving frame variables is decomposed on the above Catan forms
\begin{eqnarray}
\label{d(uv)=}
d_{u,v}=
 \Omega^{(0)} {\bb D}^{(0)}+
{1\over 2} \Omega^{ij}{\bb D}^{ij} + \Omega^{=i}{\bb D}^{\# i}+ \Omega^{\# i}{\bb D}^{= i}
\; .  \qquad
\end{eqnarray}
This decomposition defines the 'covariant harmonic derivatives' ({\it cf.} \cite{GIKOS}) which obey the Lorentz group algebra
\footnote{One can write the general expression $d_{u,v}= {1\over 2} \Omega^{(a)(b)}{\bb D}_{(a)(b)}$ with ${\bb D}_{(a)(b)}=u_{c[(a)} {\partial \over \partial u_c^{(b)]}}$, but then the last part of Eq. (\ref{d(uv)=}) would include some strange looking coefficients, ${\bb D}^{\# i}\mapsto -{1\over 2}{\bb D}^{\# i}$, ${\bb D}^{=i}\mapsto -{1\over 2}{\bb D}^{=i}$, and this is what we would like to escape. With our choice ${\bb D}^{=i} = u_{c}^i {\partial \over \partial u_c^{\#}}+ {1\over 2} u_{c}^= {\partial \over \partial u_c^{i}}$, ${\bb D}^{\# i} = u_{c}^i {\partial \over \partial u_c^{=}}+ {1\over 2} u_{c}^{\#}{\partial \over \partial u_c^{i}}$. But it is much more practical to calculate the action of ${\bb D}^{=i}$ and ${\bb D}^{\# i}$ on functions of moving frame and spinor moving frame variables  by using Eqs. (\ref{Du--=})--(\ref{d(uv)=}).}.
Using the above explicit expressions one can easily find their action on the moving frame variables,
\begin{eqnarray}
\label{D0(u)=}
  {\bb D}^{(0)}u_a^{\#}= 2 u_a^{\#} \; , \qquad  & {\bb D}^{(0)}u_a^{=}= -2 u_a^{=} \; , \qquad & {\bb D}^{(0)}u_a^{i}= 0 \; , \qquad      \nonumber
\\ \label{Dij(u)=}
{\bb D}^{ij} u_a^{\#}= 0 \; , \qquad & {\bb D}^{ij} u_a^{=}= 0 \; , \qquad & {\bb D}^{ij} u_a^{k}= 2\delta^{k[i}  u_a^{j]} \; , \qquad \nonumber
\\ \label{D++i(u)=}
{\bb D}^{\#  i }u_a^{\#}= 0 \; , \qquad & {\bb D}^{\# i}u_a^{=} = u_a^{i}
\; , \qquad &  {\bb D}^{\# i}u_a^{j} =  {1\over 2} u_a^{\#} \delta^{ij}
\; , \qquad  \nonumber
\\ \label{D--i(u)=}
{\bb D}^{=  i }u_a^{\#}= u_a^{i} \; , \qquad & {\bb D}^{= i}u_a^{=} = 0
\; , \qquad & {\bb D}^{= i}u_a^{j} = {1\over 2} u_a^{=} \delta^{ij}
\; ,  \qquad
\end{eqnarray}
and on the spinor moving frame variables,
\begin{eqnarray}
\label{D0(uv)=}
& {\bb D}^{(0)} v_{\dot{q}}^{+\alpha}= v_{\dot{q}}^{+\alpha}\; , \quad  {\bb D}^{(0)}v_{{q}}^{-\alpha}=-v_{{q}}^{-\alpha}\; , & \qquad \\ & \label{Dij(uv)=}
{\bb D}^{ij} v_{\dot{q}}^{+\alpha}= {1\over 2}
v_{\dot{p}}^{+\alpha}\gamma_{\dot{p}\dot{q}}^{ij} \; , \quad  {\bb D}^{ij}v_{{q}}^{-\alpha}={1\over 2}
v_p^{-\alpha}\gamma_{pq}^{ij}\; , & \qquad
\\ \label{D++i(uv)=}
& {\bb D}^{\#  i } v_{\dot{q}}^{+\alpha}= 0 \; , \quad  {\bb D}^{\# i}v_{{q}}^{-\alpha}=
- {1\over 2} \gamma_{q\dot{q}}^{i}v_{\dot{q}}^{+\alpha} \; , & \qquad \\ & \label{D--i(uv)=}
{\bb D}^{=  i } v_{\dot{q}}^{+\alpha}= - {1\over 2}v_{{q}}^{-\alpha}  \gamma_{q\dot{q}}^{i}\; , \quad  {\bb D}^{= i}v_{{q}}^{-\alpha}=0\; . & \qquad
\end{eqnarray}

When we discuss the spinor moving frame fields depending on worldsheet coordinates, we can introduce their covariant momentum, whcih can be denoted by the same symbols   ${\bb D}^{\#  i }$, ${\bb D}^{= i }$, ${\bb D}^{ij }$ and ${\bb D}^{(0)}$ and obey the straightforward OPE generalization of Eqs. (\ref{D++i(u)=}) and (\ref{D++i(uv)=}).

\section{Pure spinor path integral measure and spinor moving frame}
\label{d11l=sec}

The straightforward way to relate the spinor moving frame formulation of superstring \cite{BZ-str} with the pure spinor approach of \cite{Berkovits:2000fe}--\cite{Berkovits:2006vi} is to perform the covariant quantization of the former and to search for the place were the complexification, characteristic for the pure spinor approach \cite{Berkovits:2000fe}--\cite{Berkovits:2006vi}, and the  pure spinor constraint appear at the stage of regularization. For M0-brane, this is to say 11D massless superparticle, this program was completed in  \cite{IB07:M0}. However, for the superstring this way seems  to be very difficult technically. Thus here we will be rather trying to apply a more practical and intuitive approach, which was initiated by W. Siegel in 80-th \cite{Siegel:1985xj} and is used as a basic in pure spinor formalism.

Namely, we will assume that the spinor moving frame variables and their covariant momentum (${\bb D}^{\# i}$ etc.)  enter a free CFT  action for  superstring, similar to (\ref{Smin=}) or (\ref{Snmin=}), and use them   to construct the objects characteristic for pure spinor formalism and the measure of the path integral describing the loop amplitudes of the superstring.

\subsection{Pure spinor measure $d^{11}{\lambda}$ and  holomorphic 4 form on ${\bb S}^8$ space of spinor moving frame variables}\label{d11l=subsec}

Let us return to the general solution of the $D=10$ pure spinor constraint (\ref{lsl=0}) \cite{IB07:M0},
\begin{eqnarray}\label{l=L+v-2}
\lambda^\alpha =\Lambda^{+}_q v_q^{-\alpha} \; , \qquad \Lambda^{+}_q\Lambda^{+}_q=0 \; . \qquad
\end{eqnarray}
In addition to the spinor moving frame variable $v_q^{-\alpha}$, which can be considered as  homogeneous coordinates of the coset $Spin(1,9)/[SO(1,1)\otimes SO(8)\subset\!\!\!\!\!\!\times K_8]={\bb S}^8$,   and, thus, carries just $8$ degrees of freedom (d.o.f.-s), Eq. (\ref{l=L+v-}) involves the eight-component  complex light-like vector $\Lambda^{+}_q$. This latter carries 7 complex or 14 real d.o.f.-s, thus completing the real dimension 8 of $S^8$  till $22$, which is equivalent to $11$ complex dimension of the space of  $D=10$ pure spinor. Furthermore, the complex null--vector $\Lambda^{+}_q$ carries the ghost number +1, characteristic for the pure spinor $\lambda^\alpha$, and  can be related with the ghost for the $\kappa$-symmetry of superstring (which is irreducible in the spinor moving frame formulation \cite{BZ-str})\footnote{The relation between complex null vector $\Lambda^{+}_q$ and real bosonic ghost for $\kappa$--symmetry
was studied for 11D massless superparticle in \cite{IB07:M0} (there $q=1,...,16$); the relation for the case of D=10 superstring  should be similar, although the work on  establishing it promises to be quite  involving  technically.}.

The expression for the  measure $d^{11}\lambda$ in Eq. (\ref{d11L=Berk}) involves  the 'conjugate' pure spinor $\bar{\lambda}_\alpha$ which is restricted by $\lambda\bar{\lambda}\not=0$ and the pure spinor constraint (\ref{bltsbl=0}). This can be solved in terms of complimentary spinor moving frame variable
$v_{\alpha q}{}^+$ and another complex vector $\tilde{\Lambda}^{-}_q$,
\begin{eqnarray}\label{bl=tL-v+}
\bar{\lambda}_\alpha =\tilde{\Lambda}{}^{-}_q v_{\alpha q}{}^+ \; , \qquad \tilde{\Lambda}{}^{-}_q\tilde{\Lambda}{}^{-}_q=0 \; . \qquad
\end{eqnarray}
This latter  is restricted by the condition $\tilde{\Lambda}{}^-_q\Lambda^{+}_q\not= 0$.
This suggests  to write it in the form $\tilde{\Lambda}{}^{-}_q =\tilde{\Lambda}{\Lambda}{}^{-}_q $
where $\Lambda^{-}_q\Lambda^{+}_q=1$ and $\tilde{\Lambda}\not= 0$ is equal to the  product of the pure spinor and conjugate pure spinor, $\lambda\bar{\lambda}= \tilde{\Lambda}$. To resume,
\begin{eqnarray}\label{bl=tLL-v+}
\bar{\lambda}_\alpha &=&\tilde{\Lambda}{}{\Lambda}{}^{-}_q v_{\alpha q}{}^+ \; , \qquad \\ \label{L+L-=1} && {\Lambda}{}^{-}_q{\Lambda}{}^{-}_q=0 \; , \qquad {\Lambda}{}^{-}_q{\Lambda}{}^{+}_q=1 \; . \qquad
\end{eqnarray}
An equivalent form of Eq. (\ref{bl=tLL-v+}) is
\begin{eqnarray}\label{bl:lbl=}
{\bar{\lambda}_\alpha\over {\lambda}\bar{\lambda}} ={\Lambda}{}^{-}_q v_{\alpha q}{}^+ \;  . \qquad
\end{eqnarray}
Then, using the properties of spinor moving frame variables (see Eqs. (\ref{Udec}), (\ref{v+sv+=u++})--(\ref{v-sv+=ui})
(\ref{V-1V=ISTR})), we find
\begin{eqnarray}\label{bls:lbl=}
{(\bar{\lambda}\tilde{\sigma}{}^a)^\alpha\over {\lambda}\bar{\lambda}} = u^{a\#}  {\Lambda}{}^{-}_q v^{-\alpha}_{q}+ u^{a i}  {\Lambda}{}^{-}_q \gamma^{i}_{q\dot{q}}v^{+\alpha}_{\dot{q}}\; .  \qquad
\end{eqnarray}
Furthermore, after some algebra, we obtain the following expression for the algebraic factor in the pure spinor measure (\ref{d11L=Berk}):
\begin{eqnarray}\label{bls3:lbl3=}
{(\bar{\lambda}\tilde{\sigma}{}^a)^{[\alpha_1}\, (\bar{\lambda}\tilde{\sigma}{}^b)^{\alpha_2}(\bar{\lambda}\tilde{\sigma}{}^c)^{\alpha_3} \tilde{\sigma}_{abc}^{\alpha_4\alpha_5]} \over ({\lambda}\bar{\lambda})^3} \; & = 2 v_q^{-[\alpha_1}v_{\dot{q}_2}^{+\alpha_2}v_{\dot{q}_3}^{+\alpha_3}v_{\dot{q}_4}^{+\alpha_4}v_{\dot{q}_5}^{+\alpha_5]}
& \left( \gamma^{ijk}_{q\dot{q}_2}   ({\Lambda}^-\gamma^{i})_{\dot{q}_3}({\Lambda}^-\gamma^{j})_{\dot{q}_4} ({\Lambda}^-\gamma^{k})_{\dot{q}_5}- \right.   \qquad \nonumber \\ && \left. \qquad -
3{\Lambda}^-_q ({\Lambda}^-\gamma^{i})_{\dot{q}_2} ({\Lambda}^-\gamma^{j})_{\dot{q}_3} \gamma^{ij}_{\dot{q}_4\dot{q}_5}
\right)
\qquad
\end{eqnarray}
Now,  as the indices $q$ and $\dot{q}$ enumerating spinor moving frame variables take only 8 values and the indices in
$d\lambda^{\beta_1}\wedge ... \wedge d\lambda^{\beta_{11}}$ in (\ref{d11L=Berk}) are antisymmetrized with the five ones in (\ref{bls3:lbl3=}), one concludes that, after decomposing $d\lambda^{\beta}$ on $v_q^{-\beta}$ and $v_{\dot{q}_2}^{+\beta}$ (see Eq. (\ref{Dv-1-q=})), only the term $\propto v_{q_1}^{-[\beta_1 }\ldots v_{q_7}^{-\beta_7 }v_{\dot{q}_1}^{+\beta_8}\ldots v_{\dot{q}_1}^{+\beta_{11}]}$ in $d\lambda^{\beta_1}\wedge ... \wedge d\lambda^{\beta_{11}}$ contributes to $d^{11}\lambda$.

Furthermore, as far as, according to (\ref{l=L+v-}),
\begin{eqnarray}\label{dl=dL+v-+}
d\lambda^\alpha =d\Lambda^{+}_q v_q^{-\alpha} + \Lambda^{+}_q dv_q^{-\alpha}
= D\Lambda^{+}_q v_q^{-\alpha} - {1\over 2} \Omega^{=i} \Lambda^{+}_q \gamma^i_{q\dot{q}}v_{\dot{q}}^{+\alpha}
\; , \qquad
\end{eqnarray}
where $D$ are $SO(1,1)\times SO(8)$ covariant derivatives defined in (\ref{D++i(u)=}) and (\ref{D++i(uv)=}), we find that the nonvanishing contribution from $ d\lambda^{\beta_1}\wedge ...  \wedge d\lambda^{\beta_{11}}$ to the pure spinor measure should be proportional to $D\Lambda^+_{q_1}\wedge \ldots \wedge D\Lambda^+_{q_7}$ and to $\Omega^{= i_1}\wedge \ldots \wedge \Omega^{= i_4}(\Lambda^+\gamma^{i_1}_{\dot{q_1}})\ldots (\Lambda^+\gamma^{i_4}_{\dot{q_4}}) $. Hence we can rewrite the original pure spinor measure (\ref{d11L=Berk}) in the form of
\begin{eqnarray}\label{d11L=7lwOm4}
d^{11}\lambda = \propto \epsilon^{qq_1\ldots q_7} D\Lambda^+_{q_1}\wedge \ldots \wedge D\Lambda^+_{q_7}
\wedge \Omega^{= i_1}\wedge \ldots \wedge \Omega^{= i_4} \; \epsilon^{\dot{q}_1\ldots \dot{q}_8}
(\Lambda^+\gamma^{i_1}_{\dot{q_1}})\ldots (\Lambda^+\gamma^{i_4}_{\dot{q_4}}) \times \nonumber \\
\times
 \left( \gamma^{ijk}_{q\dot{q}_5}   ({\Lambda}^-\gamma^{i})_{\dot{q}_6}({\Lambda}^-\gamma^{j})_{\dot{q}_7} ({\Lambda}^-\gamma^{k})_{\dot{q}_8}-
3{\Lambda}^-_q ({\Lambda}^-\gamma^{i})_{\dot{q}_5} ({\Lambda}^-\gamma^{j})_{\dot{q}_6} \gamma^{ij}_{\dot{q}_7\dot{q}_8}
\right)
\; . \qquad
\end{eqnarray}
We intentionally have not fixed the inessential numerical coefficient in the above expression for the pure spinor measure and have written it with $\propto$ symbol, because this  expression  happens to be intermediate. Below we will change it by expressing  $D\Lambda^+_{q}$ factors in terms of $SO(8,{\bb C})$ Cartan forms which we are going to define in sec. \ref{SO8=CF} after introducing,  in the next sec. \ref{r=sol}, a set of variables parametrizing the $SO(8,{\bb C})$ group and using them, together with spinor moving frame, to solve the constraints on the fermionic spinor $r_\alpha$ of the nonminimal pure spinor formalism.

\subsection{SO(8,C) frame and solution of the constraints for $r_\alpha$}\label{r=sol}

One can complete the set of complex null vectors $\Lambda^{\pm}_q$ till the complete basis in the space of 8-vectors by introduced the set of 6 mutually orthogonal and normalized vectors $\Lambda_q^I$, which are orthogonal to  $\Lambda^{\pm}_q$,
\begin{eqnarray}\label{L+L-LI}
{\Lambda}{}^{-}_q{\Lambda}{}^{-}_q=0 \; , \qquad {\Lambda}{}^{+}_q{\Lambda}{}^{+}_q=0 \; , \qquad {\Lambda}{}^{-}_q{\Lambda}{}^{+}_q=1 \; , \qquad \nonumber \\
{\Lambda}{}^{\pm}_q{\Lambda}{}^{I}_q=0 \; , \qquad {\Lambda}{}^{I}_q{\Lambda}{}^{J}_q=\delta^{IJ} \; . \qquad
\end{eqnarray}
Such a set of vectors can be collected in an $[SO(8)]^c=SO(8,C)$ valued matrix
\begin{eqnarray}\label{LinSO}
{\Lambda}{}^{(p)}_q= \left({\Lambda}{}^{I}_q\, , \, {1\over 2}({\Lambda}{}^{+}_q + {\Lambda}{}^{-}_q) \, , \, {i\over 2}({\Lambda}{}^{+}_q - {\Lambda}{}^{-}_q) \right) \quad \in \quad SO(8,{C})\; . \qquad
\end{eqnarray}
Eqs. (\ref{L+L-LI}) then appear as ${\Lambda}{}^{(p)}_q{\Lambda}{}^{(p^\prime)}_q=\delta{}^{pp^\prime}$,
which is an equivalent form of  (\ref{LinSO}). Its another equivalent form,  ${\Lambda}{}^{(p)}_q{\Lambda}{}^{(p)}_{q^\prime}=\delta_{qq^\prime}$, gives rise to the completeness condition
\begin{eqnarray}\label{I=SO}
\delta_{qp}= {\Lambda}{}^{+}_q {\Lambda}{}^{-}_p+ {\Lambda}{}^{-}_q {\Lambda}{}^{+}_p + {\Lambda}{}^{I}_q {\Lambda}{}^{I}_p\; . \qquad
\end{eqnarray}

The additional vectors ${\Lambda}{}^{I}_q$ are  useful, in particular,  to solve the constraints (\ref{bltsr=0}) imposed on the fermionic spinor $r_\alpha$,
\begin{eqnarray}\label{rf=}
r_\alpha= \, \chi \, {\Lambda}{}^{-}_q   v_{\alpha q}^+ + \, \chi^I \,  {\Lambda}{}^{I}_q  v_{\alpha q}^+ + \, \chi^i \,  {\Lambda}{}^{-}_q \gamma^i_{q\dot{q}} v_{\alpha \dot{q}}^- \; . \qquad
\end{eqnarray}
Here  the complex fermionic scalar $\chi$ and the complex fermionic $SO(6)$ vector $\chi^I:= \chi^{-I}$ are unconstrained, while
the complex fermionic SO(8) vector $\chi^i=\chi^{++i}$, with eight component, is defined up to the local transformations
\begin{eqnarray}\label{chii=chii+}
\chi^i \; \sim \; \chi^i + \Lambda^-_q \gamma^i_{q\dot{q}} \chi^{(1)}_{\dot{q}}\;  \qquad
\end{eqnarray}
with the 8-component parameter $\chi^{(1)}_{\dot{q}}:=\chi^{(1)+++}_{\dot{q}}$; furthermore, this latter is defined up to the transformations
\begin{eqnarray}\label{chidq=chdq+}
 \chi^{(1)}_{\dot{q}} \; \sim \;  \chi^{(1)}_{\dot{q}} + \Lambda^-_q \gamma^i_{q\dot{q}} \chi^{(2)i}\;  \qquad
\end{eqnarray}
with 8-component parameter $\chi^{(2)i}:=\chi^{(2)i++++}$ defined, in its turn, up to the transformations
\begin{eqnarray}\label{chii=chii+2}
\chi^{(2)i} \; \sim \; \chi^{(2)i}+ \Lambda^-_q \gamma^i_{q\dot{q}} \chi^{(3)}_{\dot{q}}\; ,  \qquad \ldots
\end{eqnarray}

Multidots in (\ref{chii=chii+}) mark that this process of finding 8-parametric indefiniteness of the parameters of symmetry can be continued up to infinity so that, like in the case of $\kappa$--symmetry of the standard Green--Schwarz formulation of superstring \cite{G+S84}, we are dealing with the infinitely reducible symmetry. This implies that the effective number of the parameters  $\chi^i$ in (\ref{rf=}), which is reduced by the above gauge symmetry, has to be calculated as an infinite sum $8-8+8-...$. As usually, regularizing this expression by identifying it with the limit of geometric progression, we find
$8-8+8-...= 8\, \lim\limits_{q\mapsto 1}(1-q+q^2-...)= 8\, \lim\limits_{q\mapsto 1}{1\over 1+q}= 4$. This completes the number of component of $\chi^I$ and $\chi$, 6+1=7, till  $11$, which is the number of components of the spinor $r_\alpha$ constrained by (\ref{bltsr=0}) \cite{Berkovits:2005bt}. This simple calculation gives an evidence that our (\ref{rf=}) is the general solution of (\ref{bltsr=0}).

\subsection{SO(8,C) Cartan forms and the pure spinor measure $d^{11}{\lambda}$} \label{SO8=CF}

Furthermore, the above group theoretical interpretation  of the basis involving the complex light-like vectors ${\Lambda}{}^{+}_q$ and ${\Lambda}{}^{-}_q$ allows to define in an easy way the derivatives with respect to this constrained variables and also the measure with respect to them.

Let us introduce the set of  Cartan forms  including $SO(6,\mathbf{C})$ and $SO(2,\mathbf{C})$ connections
\begin{eqnarray}\label{OmLIJ=}
\Omega_{\Lambda}^{IJ}:= {\Lambda}{}^{I}_q d{\Lambda}{}^{J}_q\; , \qquad \Omega_{\Lambda}^{(0)}= {\Lambda}{}^{-}_q d{\Lambda}{}^{+}_q
\; , \qquad
\end{eqnarray}
as well as $SO(8,C)/[SO(8,C)\otimes SO(2,\mathbf{C})]$ vielbein forms
\begin{eqnarray}\label{OmL+I=}
\Omega_{\Lambda}^{\pm J}:= {\Lambda}{}^{\pm}_q d{\Lambda}{}^{J}_q
\; . \qquad
\end{eqnarray}
[The subindex $\Lambda$ is introduced to make a difference with (real) $SO(1,9)$ Cartan forms in Eqs. (\ref{Du--=})-- (\ref{Dv-1+dq=})].

The derivatives of the $\Lambda^{\pm}_q$ and $\Lambda^{I}_q$ vectors can be expressed as
\begin{eqnarray}\label{dL+q=}
d{\Lambda}{}^{ +}_q = \; {\Lambda}{}^{ +}_q \Omega_{\Lambda}^{(0)} - {\Lambda}{}^{I}_q \Omega_{\Lambda}^{+I}\; ,
\qquad
\\ \label{dL-q=} d{\Lambda}{}^{-}_q = - {\Lambda}{}^{ -}_q \Omega_{\Lambda}^{(0)} - {\Lambda}{}^{I}_q \Omega_{\Lambda}^{-I}\; , \qquad
\\ \label{dLIq=} d{\Lambda}{}^{I}_q = {\Lambda}{}^{ +}_q   \Omega_{\Lambda}^{-I}+{\Lambda}{}^{ -}_q   \Omega_{\Lambda}^{+I}- {\Lambda}{}^{J}_q \Omega_{\Lambda}^{JI}
\; . \qquad
\end{eqnarray}

Decomposing the exterior derivative in  the $SO(8,C)$ group manifold (i.e. in  the space parametrized by  $\Lambda^{\pm}_q$ and $\Lambda^{I}_q$) on  the Cartan forms,
\begin{eqnarray}\label{dL=OmDL}
d_{\Lambda }= \Omega^{+J}_\Lambda D^{-J}_\Lambda+ \Omega^{-J}_\Lambda D^{+J}_\Lambda + \Omega^{(0)}_\Lambda D^{(0)}_\Lambda + {1\over 2} \Omega^{IJ}_\Lambda D^{IJ}_\Lambda \, ,
\;  \qquad
\end{eqnarray}
we obtain the covariant derivative $D^{-J}_\Lambda$, $D^{(0)}_\Lambda$ $D^{IJ}_\Lambda$ generating the SO(8) algebra. This statement can be easily checked using their action on the basic variables $\Lambda^{\pm}_q$ and $\Lambda^{I}_q$,
\begin{eqnarray}\label{DL0DLIJ}
D^{(0)}_\Lambda \Lambda^{\pm}_q = \pm \Lambda^{\pm}_q \; , \qquad D^{(0)}_\Lambda \Lambda^{I}_q =0 \; , \qquad D^{IJ}_\Lambda \Lambda^{\pm}_q = 0 \; , \qquad D^{IJ}_\Lambda \Lambda^{K}_q = 2 \Lambda^{[I}_q \delta^{J]K}\; , \qquad \\ \label{DL-IL=}
D^{-I}_\Lambda \Lambda^{+}_q =- \Lambda^{I}_q \; , \qquad D^{-I}_\Lambda \Lambda^{-}_q =0 \; , \qquad D^{-I}_\Lambda \Lambda^{J}_q = \Lambda^{-}_q \delta^{IJ}\; , \qquad \\ \label{DL+IL=}
D^{+I}_\Lambda \Lambda^{+}_q =0 \; , \qquad D^{+I}_\Lambda \Lambda^{-}_q = - \Lambda^{I}_q \; , \qquad D^{+I}_\Lambda \Lambda^{J}_q = \Lambda^{+}_q \delta^{IJ}\; .  \qquad
\end{eqnarray}

Clearly, to use our formalism in covariant description of quantum superstring, we need rather to define some covariant derivatives, i.e. derivative covariant under $ SO(8)\times SO(1,1)$.

Notice that only $SO(8,\mathbf{R})\subset SO(8,\mathbf{C})$  and $SO(1,1)\subset SO(2,\mathbf{C}))$ are symmetries of our construction, while $U(1)\subset SO(2,\mathbf{C})$ is the phase rotation of pure spinors and, to preserve (\ref{l=L+v-}) under  $SO(8,\mathbf{C})/SO(8,\mathbf{R})$ transformations of $\Lambda^+_q$, one should need to multiply the real spinor moving frame field $v^{-\alpha}_q$ by an unavoidably complex
matrix\footnote{In this respect it is interesting that, in a recent \cite{Movshev:2012},  Movshev has begun to develop in 11D the line similar to our basic construction in \cite{IB07:M0}, but with the  complexification of the 11D Lorentz group $SO(1,10)$. The aim in \cite{Movshev:2012} is to construct an 11D twistor
transform related to the proposition of Cederwall \cite{Cederwall=11D} for the
off--shell action of 11D supergravity in 11D pure spinor superspace. We should also notice that, to our best knowledge, for the first time the twistor transform in D=11 was discussed in \cite{GHT93}, although without relation to pure spinors.}.
Due to this reason, it is useful to define counterparts $\Xi^{(0)}$, $\Xi^{IJ}$, $\Xi^{\pm I}$ of the above Cartan forms $\Omega_\Lambda^{(0)}$, $\Omega_\Lambda^{IJ}$, $\Omega_\Lambda^{\pm I}$, that are related with the $SO(8,R)\otimes SO(1,1)\otimes SO(6,C)$ covariant derivatives of the complex vectors $\Lambda^{\pm}_q$ and $\Lambda^{I}_q$ (notice that, actually,  $D{\Lambda}{}^{ +}_q$ has already appeared in Eq. (\ref{dl=dL+v-+})),
\begin{eqnarray}\label{DL+q=}
D{\Lambda}{}^{ +}_q :=& d{\Lambda}{}^{ +}_q - {\Lambda}{}^{ +}_q \Omega^{(0)}-
 {\Lambda}{}^{ +}_p \Omega_{pq} &=: \; {\Lambda}{}^{ +}_q \Xi^{(0)}  - {\Lambda}{}^{I}_q \Xi^{+I}
 \; ,
\qquad
\\ \label{DL-q=} D{\Lambda}{}^{-}_q :=& d{\Lambda}{}^{-}_q + {\Lambda}{}^{ -}_q \Omega^{(0)}-
 {\Lambda}{}^{ -}_p \Omega_{pq}   &=: - {\Lambda}{}^{ -}_q \Xi^{(0)} - {\Lambda}{}^{I}_q \Xi^{-I}\; , \qquad
\\ \label{DLIq=}D{\Lambda}{}^{I}_q :=& d{\Lambda}{}^{I}_q  + {\Lambda}{}^{ J}_q   \Omega_{\Lambda}^{JI}- {\Lambda}{}^{I}_p \Omega_{pq} &=: \; \Lambda^+_q \Xi^{-I}+  \Lambda^-_q \Xi^{+I} + {\Lambda}{}^{J}_q   \Xi^{JI}
\; . \qquad
\end{eqnarray}
These are
\begin{eqnarray}\label{Xi0=}
\Xi^{(0)} = \Omega_{\Lambda}^{(0)} - \Omega^{(0)} - {\Lambda}{}^{ +}\Omega{\Lambda}{}^{ -}\; ,  \qquad  & with &  \qquad {\Lambda}{}^{ +}\Omega{\Lambda}{}^{-}:= {\Lambda}{}^{ +}_q\Omega_{pq}{\Lambda}{}^{-}_p \; \qquad \\
\label{Xi+I=}
\Xi^{+I} = \Omega_{\Lambda}^{+I} + {\Lambda}{}^{ +}\Omega{\Lambda}{}^{I}\; ,  \qquad & with & \qquad {\Lambda}{}^{ +}\Omega{\Lambda}{}^{I}:= {\Lambda}{}^{ +}_p\Omega_{pq}{\Lambda}{}^{ I}_q\; \qquad \\
\label{Xi-I=}
\Xi^{-I} = \Omega_{\Lambda}^{-I} + {\Lambda}{}^{ -}\Omega{\Lambda}{}^{I}\; ,  \qquad  & with &  \qquad {\Lambda}{}^{ -}\Omega{\Lambda}{}^{I}:= {\Lambda}{}^{ -}_p\Omega_{pq}{\Lambda}{}^{ I}_q\; \qquad \\
\label{XiIJ=}
\Xi^{IJ} =  \Omega_{\Lambda}^{IJ} + {\Lambda}{}^{ I}\Omega{\Lambda}{}^{J}\; ,  \qquad  & with & \qquad {\Lambda}{}^{ I}\Omega{\Lambda}{}^{J}:= {\Lambda}{}^{ I}_p\Omega_{pq}{\Lambda}{}^{J}_q
\end{eqnarray}
In particular, Eq. (\ref{DL+q=}) allows to express the derivative of the composed pure spinor (\ref{l=L+v-}) as
\begin{eqnarray}\label{dl=Om+Xi}
d\lambda^\alpha = -{1\over 2} \Omega^{=i}\, \Lambda^+_q\gamma^i_{q\dot{q}}v^{+\alpha}_{\dot{q}} +
(\Lambda^+_q\Xi^{(0)}  - {\Lambda}{}^{I}_q \Xi^{+I}) v^{-\alpha}_{{q}}
\; . \qquad
\end{eqnarray}
Notice also that, as far as we defined $\tilde{\Lambda}= \lambda^\alpha\bar{\lambda}_\alpha$,  using (\ref{bl=tL-v+}) we find
\begin{eqnarray}\label{dlbl=blXi}
d\lambda^\alpha\,  \bar{\lambda}_\alpha= d\tilde{\Lambda} - \lambda^\alpha d\bar{\lambda}_\alpha=
 \tilde{\Lambda} \Xi^{(0)}
\; . \qquad
\end{eqnarray}

Now, using (\ref{DL+q=}) we can write the pure spinor measure (\ref{d11L=7lwOm4}) in the form
\begin{eqnarray}\label{d11L=hOm}
\fbox{$\, d^{11}\lambda = \epsilon^{I_1\ldots I_6} \Xi^{(0)} \wedge \Xi^{+I_1}\wedge \ldots \wedge \Xi^{+I_6}
\wedge \Omega^{= i_1}\wedge \ldots \wedge \Omega^{= i_4} \; \; M^{++ i_1...i_4}(\Lambda^\pm)
\, $} \; , \qquad
\end{eqnarray}
where
\begin{eqnarray}\label{M++iiii=MeM}
&& M^{++ i_1...i_4}(\Lambda^\pm):= (\Lambda^+\gamma^{i_1})_{\dot{q_1}}\ldots (\Lambda^+\gamma^{i_4})_{\dot{q_4}} \; \epsilon^{\dot{q}_1\ldots \dot{q}_4\dot{p}_1\ldots \dot{p}_4} \; M^{-- }_{\dot{p}_1\ldots \dot{p}_4}(\Lambda^\pm)\; , \qquad  \;  .
\end{eqnarray}
with
\begin{eqnarray}
 \label{M--=} && M^{-- }_{\dot{p}_1\ldots \dot{p}_4}(\Lambda^\pm) :=
({\Lambda}^+\gamma^{ijk})_{\dot{p}_1} ({\Lambda}^-\gamma^{i})_{\dot{p}_2}({\Lambda}^-\gamma^{j})_{\dot{p}_3} ({\Lambda}^-\gamma^{k})_{\dot{p}_4} - 3({\Lambda}^-\gamma^{i})_{\dot{p}_1}({\Lambda}^-\gamma^{j})_{\dot{p}_2} \gamma^{ij}_{\dot{p}_3\dot{p}_4}
\;  .
\end{eqnarray}
In (\ref{d11L=7lwOm4}) all the 11 directions of the  integration are represented by the generalized Cartan forms, seven of which correspond to a coset of $SO(8,C)$ and 4 to the coset of the Lorentz group $Spin(1,9)$. Let us stress that, although this latter 4 form contains wedge product   $\Omega^{= i_1}\wedge \ldots \wedge \Omega^{= i_4} $ of real vielbein of ${\bb S}^8$, these are contracted with the complex $SO(8)$ tensor $M^{++ i_1...i_4}(\Lambda^\pm)
\, $ so that the corresponding contribution    $\Omega^{= i_1}\wedge \ldots \wedge \Omega^{= i_4} \; \; M^{++ i_1...i_4}(\Lambda^\pm) \, $ can be considered as a complex, {\it holomorphic} measure. Furthermore, as far as the complex $SO(8)$ tensor $M^{++ i_1...i_4}(\Lambda^\pm)$ depends on complex null vector $\Lambda^+_q$ and its dual
$\Lambda^-_q$, that is the complex holomorphic measure  on  the fiber ${\bb S}^8$ of the space of 10D pure spinors considered as fiber bundle with the base ${\bb C}^7-\{ 0\}$.

\subsection{The `conjugate' pure spinor measure $d^{11}\bar{\lambda}$}\label{d11bL=sec}

Let us turn to the measure for the 'conjugate' pure spinor of the nonminimal formalism, Eq. (\ref{d11bL=}). Its derivative  reads
 \begin{eqnarray}\label{dbl=Om+Xi}
d\bar{\lambda}_\alpha = {1\over 2} \Omega^{\# i}\, \tilde{\Lambda} \Lambda^-_q \gamma^i_{q\dot{q}}v_{\alpha\dot{q}}{}^{\! -} + [(d\tilde{\Lambda} - \tilde{\Lambda}\Xi^{(0)})
\Lambda^-_q
 - {\Lambda}{}^{I}_q \Xi^{-I})] v_{\alpha {q}}{}^+
\; . \qquad
\end{eqnarray}
This is related to (\ref{dbl=Om+Xi}), essentially,  by reversing the $SO(1,1)$ weight, substituting the upper $Spin(1,9)$ index by lower one (thus passing from spinor moving frame variables to the variables forming the inverse spinor moving frame matrix) and replacing $\Xi^{(0)}$ by $(d\tilde{\Lambda} - \tilde{\Lambda}\Xi^{(0)})$ (see Eq. (\ref{dlbl=blXi})). The algebraic factors in
(\ref{d11L=Berk}) and (\ref{d11bL=}) are related in a similar manner so that, as far as we are not interested in an overall  numerical multiplier, we can write immediately the final answer for the 'conjugate' pure spinor measure (\ref{d11bL=}). It reads
 \begin{eqnarray}\label{d11bL=hOm}
d^{11}\bar{\lambda} =
{\tilde{\Lambda}^6\over 6!}\;  d\tilde{\Lambda} \wedge  \; \epsilon^{I_1\ldots I_6}  \Xi^{-I_1}\wedge \ldots \wedge \Xi^{-I_6}
\wedge \Omega^{\# i_1}\wedge \ldots \wedge \Omega^{\# i_4} \;   M^{-- i_1...i_4}(\Lambda^\pm)
\; , \qquad
\end{eqnarray}
with
\begin{eqnarray}\label{M--iiii=MeM}
&& M^{-- i_1...i_4}(\Lambda^\pm):= (\Lambda^-\gamma^{i_1})_{\dot{q_1}}\ldots (\Lambda^-\gamma^{i_4})_{\dot{q_4}} \; \epsilon^{\dot{q}_1\ldots \dot{q}_4\dot{p}_1\ldots \dot{p}_4} \; M^{++}_{\dot{p}_1\ldots \dot{p}_4}(\Lambda^\pm)\; , \qquad  \\ \label{M++=} && M^{++}_{\dot{p}_1\ldots \dot{p}_4}(\Lambda^\pm) :=
({\Lambda}^-\gamma^{ijk})_{\dot{p}_1} ({\Lambda}^+\gamma^{i})_{\dot{p}_2}({\Lambda}^+\gamma^{j})_{\dot{p}_3} ({\Lambda}^+\gamma^{k})_{\dot{p}_4} - 3({\Lambda}^+\gamma^{i})_{\dot{p}_1}({\Lambda}^+\gamma^{j})_{\dot{p}_2} \gamma^{ij}_{\dot{p}_3\dot{p}_4}
\;  .
\end{eqnarray}

Actually, to be more precise, one has to substitute $(d\tilde{\Lambda} - \tilde{\Lambda}\Xi^{(0)})$ for $d\tilde{\Lambda} $ in (\ref{d11bL=hOm}). However, as far as we expect to use
$d^{11}\bar{\lambda}$ only together with the original pure spinor measure (\ref{d11L=hOm}), and the term $\propto \Xi^{(0)}$ does not contribute into
$d^{11}\bar{\lambda} \wedge d^{11}\bar{\lambda}$, we prefer to write a simpler expression (\ref{d11L=hOm}) from the very beginning.

Notice that the complete measure of the nonminimal pure spinor formalism includes $d^{11} {\lambda} \wedge d^{11}\bar{\lambda}$ and hence contains both $\Omega^{=i}$ and $\Omega^{\# i}$.  These enters inside the 8 form
\begin{eqnarray}\label{d11bL=hOm}
\Omega^{= i_1}\wedge \ldots \wedge \Omega^{= i_4}
\wedge \Omega^{\# j_1}\wedge \ldots \wedge \Omega^{\# j_4} \;   M^{++ i_1...i_4}(\Lambda^\pm) \;   M^{-- i_1...i_4}(\Lambda^\pm)
\;   \qquad
\end{eqnarray}
which provides a kind of holomorphic measure on the noncompact coset $SO(1,9)/[SO(1,1)\otimes SO(8)]$ considered as a fiber of a bundle over $SO(8,{\bb C})/[U(1)\otimes SO(6,{\bb C})]$ coset. Notice the difference with the  case of minimal pure spinor formalism where, as discussed above, the measure includes only $\Omega^{= i}$ forms entering inside
$\Omega^{= i_1}\wedge \ldots \wedge \Omega^{= i_4}M^{++ i_1...i_4}(\Lambda^\pm)$ which provides the holomorphic measure on the compact fiber ${\bb S}^8$ of the fiber bundle over ${\bb C}^7-\{ 0\}$.

\section{A problem with the fermionic measure $d^{11}r$ and a possible wayout}\label{d11r=problem}

It is natural to try to reproduce as well the measure (\ref{d11bL=}) for the constrained  fermionic variable $r_\alpha$. However, the available solution for the constraints imposed on this variable, Eq. (\ref{rf=}), includes $\chi^i$ defined up to an infinitely reducible symmetry transformations (\ref{chii=chii+}), (\ref{chidq=chdq+}), (\ref{chii=chii+2}) (resembling the $\kappa$--symmetry of the original Green--Schwarz formulation of the superstring). This suggests to search for an alternative way to write the counterpart of the fermionic measure $d^{11}r$ in the nonminimal pure spinor path integral for the quantum superstrings.

To this end let us first concentrate on the contribution of  the constrained fermionic spinor $r_\alpha$ into the BRST charge of the nonminimal spinor formalism, Eq. (\ref{Q=lD+rw}). It enters in the product with the momentum of the conjugate pure spinor, $\bar{w}{}^\alpha r_\alpha$, and the constraints for $r_\alpha$ comes from the requirement of the preservation of the $\bar{\Xi}^a$ gauge symmetry (\ref{vbw=}) acting on   $\bar{w}{}^\alpha$.

Now, the momentum  $\bar{w}{}^\alpha$ enters also the free CFT action (\ref{Snmin=}), in
this case in the product with $\bar{\partial} \bar{\lambda}_\alpha $. At this stage let us notice that, with our solution of the pure spinor constraints (\ref{bl=tLL-v+}), $d\bar{\lambda}_\alpha = dz \partial \bar{\lambda}_\alpha+ d\bar{z} \bar{\partial} \bar{\lambda}_\alpha$ has the form (\ref{dbl=Om+Xi}) and, hence,
\begin{eqnarray}\label{bdbl=Om+Xi}
\bar{\partial} \bar{\lambda}_\alpha = {1\over 2} \Omega_{\bar{z}}^{\# i}\, \tilde{\Lambda} \Lambda^-_q \gamma^i_{q\dot{q}}v_{\alpha\dot{q}}{}^{\! -} + [(\bar{\partial}\tilde{\Lambda} - \tilde{\Lambda}\Xi_{\bar{z}}^{(0)})
\Lambda^-_q
 - {\Lambda}{}^{I}_q \Xi_{\bar{z}}^{-I})] v_{\alpha {q}}{}^+
\; . \qquad
\end{eqnarray}
Here  $\Omega_{\bar{z}}^{\# i}$ appear in the decomposition $\Omega^{\# i}=d{z}\Omega_{{z}}^{\# i}+ d\bar{z}\Omega_{\bar{z}}^{\# i}$ and, similarly,
$\Xi^{-I}=dz\Xi_{{z}}^{-I}+d\bar{z}\Xi_{\bar{z}}^{-I}$, $\Xi^{(0)}=dz\Xi_{{z}}^{(0)}+d\bar{z}\Xi_{\bar{z}}^{(0)}$ {\it etc.}.

Eq. (\ref{bdbl=Om+Xi}) implies that  the free CFT action (\ref{Snmin=}) contains $\bar{w}^\alpha$ only in the following combinations (as far as only these contribute to
$\bar{\partial} \bar{\lambda}_\alpha \bar{w}^\alpha$)
\begin{eqnarray}\label{bw=eff}
\Lambda^-_q \gamma^i_{q\dot{q}} \, \bar{w}^\alpha v_{\alpha\dot{q}}{}^{\! -} \; , \qquad
\bar{w}^\alpha v_{\alpha {q}}{}^+\, \Lambda^-_q\; , \qquad
   \bar{w}^\alpha v_{\alpha {q}}{}^+\, {\Lambda}{}^{I}_q
\; . \qquad
\end{eqnarray}
Let us observe that these combinations are invariant under the $\bar{\Xi}^a$ gauge symmetries of Eq. (\ref{vbw=}). Indeed, using the general solution (\ref{bl=tLL-v+}) one sees that under this symmetry $ \delta\bar{w}^\alpha v_{\alpha {q}}{}^+=\tilde{\bar{\Xi}}{}^{++}\Lambda^-_q$ and $\delta \bar{w}^\alpha v_{\alpha\dot{q}}{}^{\! -}=\tilde{\bar{\Xi}}{}^{i}\Lambda^-_q \gamma^i_{q\dot{q}}$, so that the variations of the expressions in  Eq. (\ref{vbw=}) vanish due to the algebraic properties of $\Lambda^\pm_q$ and $\Lambda^I_q$.

Now observe that the gauge invariant combinations in Eq. (\ref{vbw=}) are in one-to one correspondence to the 'covariant momenta', $w_{\tilde{\Lambda}}$ dual to $d\tilde{\Lambda}$  (which is to say, conjugate to $\tilde{\Lambda}$),  and
${\bb D}^{= i}$ and ${D}_{\Lambda}^{+I}$ dual to the Cartan forms of the cosets of $SO(1,9)$ and $SO(8,C)$ groups, which enter the {\it r.h.s.} of Eq. (\ref{dbl=Om+Xi}). (We denote these covariant momenta by the same symbols as the covariant derivatives in (\ref{dL=OmDL}) and (\ref{DL0DLIJ}), (\ref{DL-IL=}), (\ref{DL+IL=})).

The above observation encourages us to propose the following prescription of changing the dynamical variables and the free CFT action (presently, in its part related to conjugate pure spinor)
\begin{eqnarray}\label{bdblbw->}
\bar{\partial}\bar{\lambda}_\alpha \, \bar{w}^\alpha\quad \mapsto \quad (\bar{\partial}\tilde{\Lambda} - {\Xi}_{\bar{z}}^{(0)}  \tilde{\Lambda}) \, \bar{w}_{\tilde{\Lambda} } +{\Xi}_{\bar{z}}^{-I} {D}_{\Lambda}^{+I} + \Omega_{\bar{z}}^{\# i}
D^{= i} \; .  \qquad
\end{eqnarray}
Furthermore, we have to reformulate the nonminimal pure spinor model in such a way that it involves  $w_{\tilde{\Lambda} }$, ${D}_{\Lambda}^{+I}$ and  ${\bb D}^{= i}$  instead of
$\bar{w}^\alpha$ defined modulo $\bar{\Xi}^a$ gauge symmetry (\ref{vbw=}),
\begin{eqnarray}\label{bw->}
\bar{w}^\alpha\quad \mapsto \quad  ( \bar{w}_{\tilde{\Lambda} } , \,  {D}_{\Lambda}^{+I}, \,
{\bb D}^{=i}) \; .  \qquad
\end{eqnarray}
To check once more that Eqs. (\ref{bdblbw->}) and (\ref{bw->}) are reasonable, one can calculate the number of degrees of freedom. On the left, being a momentum conjugate to a pure spinor,   $\bar{w}^\alpha$ has to carry 11 complex degrees of freedom. On the right we have 1 complex degree of freedom in $ w_{\tilde{\Lambda} } $, 6 complex d.o.f.-s in $ {D}_{\Lambda}^{+I}$ and 4 complex (8 real) d.o.f.-s in ${\bb D}^{= i}$.

In its turn, Eqs. (\ref{bdblbw->}) and (\ref{bw->}) suggest to make the following substitution in the BRST charge of the nonminimal pure spinor approach (\ref{Q=lD+rw}):
\begin{eqnarray}\label{rbw->}
r_\alpha \, \bar{w}^\alpha\quad \mapsto \quad r^{(0)} \, \bar{w}_{\tilde{\Lambda} } +r^{-I} {D}_{\Lambda}^{+I} + r^{\# i}
{\bb D}^{=i}\;   \qquad
\end{eqnarray}
and to consider the modification of the original formalism which involves the 11 complex (22 real) fermionic fields: 1 complex  $r^{(0)}$, 6 complex in $r^{-I}$ and 8 real in $ r^{\# i}$,  instead of the constrained $r_\alpha$,
\begin{eqnarray}\label{r->}
r_\alpha \quad \mapsto \quad (r^{(0)} , \, r^{-I},\,  r^{\# i} ) \; .    \qquad
\end{eqnarray}
Then the similarity between the bosonic $d^{11}{\lambda}$ and fermionic $d^{11}r$ measures of the nonminimal pure spinor formalism, which are related by the map
$d {\lambda}{}^\alpha \mapsto {\partial \over \partial r^\alpha}$, suggests to obtain the fermionic measure $d^{11}r$  of our  approach by mapping
\begin{eqnarray}\label{d11bl->d11r}
d\Xi^{(0)} \; \mapsto {\partial \over \partial r^{(0)}}  \; ,    \quad
\Xi^{+I} \; \mapsto {\partial \over \partial r^{-I}}  \; ,    \quad
\Omega^{= i} \; \mapsto {\partial \over \partial r^{\# i}}  \; .    \qquad
\end{eqnarray}
The result is
\begin{eqnarray}\label{d11r=11M--}
d^{11}r = M^{++ i_1...i_4}(\Lambda^\pm) \epsilon^{I_1\ldots I_6} \;   {\partial \over \partial r^{-I_1}}  \; \ldots  {\partial \over \partial r^{-I_6}}  \; {\partial \over \partial r^{(0)}}    \;
\; {\partial \over \partial r^{\# i_1}}   \ldots \; {\partial \over \partial r^{\# i_4}} \;
\; , \qquad
\end{eqnarray}
where $M^{++ i_1...i_4}(\Lambda^\pm)$ is defined in Eqs. (\ref{M++iiii=MeM}) and (\ref{M--=}).

As far as non-minimal fermion contribution to the free CFT action is concerned, the natural prescription would be
\begin{eqnarray}\label{bdrs->}
s^\alpha \bar{\partial} r_\alpha \, \quad \mapsto \quad  s^{(0)}\bar{\partial} r^{(0)}  +  s^{+I}\bar{\partial} r^{-I} + s^{= i} \bar{\partial} r^{\# i}\; ,
   \qquad
\end{eqnarray}
so that
\begin{eqnarray}\label{bdrs->}
 \, s^\alpha\quad \mapsto \quad ( s^{(0)} ,\,  s^{+I} ,\,
s^{= i})\;    \qquad
\end{eqnarray}
with complex $s^{(0)}$ and $s^{+I} $, and real $s^{= i}$ (all unconstrained).

\section{Discussion and outlook. Towards quantum superstring formulation  without reducible symmetries?}
\label{discussion}

The discussion above suggests a possible reformulation of the quantum superstring theory, in particular the prescription to calculate loop superstring amplitudes reached in the frame of pure spinor approach \cite{Berkovits:2000fe}--\cite{Oda:2007ak}, in terms of new set of variables. In such a formulation the elements of the ghost sector of the nonminimal pure spinor formalism
\begin{eqnarray}\label{VARnmin:}
(\lambda^\alpha, w_\alpha )\; , \quad (\bar{w}^\alpha ,  \bar{\lambda}_\alpha) \; , \quad ( r_\alpha s^\alpha  )
\;  \qquad
\end{eqnarray}
are replaced by the constrained  complex variables $\Lambda^{\pm}_q$ parameterizing the coset of $SO(8,{\bb C})$ group (Eqs. (\ref{LinSO}) and (\ref{L+L-LI})) and its Lie algebra,  complex scalar
and its momentum $(\tilde{\Lambda}, \bar{w}_{\tilde{\Lambda}})$, and
the spinor moving frame variables parameterizing the coset $SO(1,9)/[SO(1,1)\times SO(8)]$ (Eqs. (\ref{V-1STR})--(\ref{v-sv+=ui})) and their covariant momentum
\begin{eqnarray}\label{VARnew:}
\matrix{
{SO(8,{\bb C})/ SO(6,{\bb C}) }  & =\{ (\Lambda^{+}_q\, , \Lambda^{-}_q) \}\; , \quad & \{ ({D}_{\Lambda}^{(0)},   {D}_{\Lambda}^{\pm I}) \} \;  , \cr
 {\bb C}^1-\{ 0\} &= \{ \tilde{\Lambda}\} \; , \qquad &  \{ \bar{w}_{\tilde{\Lambda}}\}\; , \cr  {SO(1,9)\over SO(1,1)\times SO(8)} &= \left\{( v^{-\alpha}_q\,  , v^{+\alpha}_q)\,  \vert_{_{mod (SO(1,1)\times SO(8))}}  \right\}\; , \quad & \{ ({\bb D}^{\# i}\, , \, {\bb D}^{=i})\} \;  .} \qquad
\end{eqnarray}
(Notice by pass that, following the line of sec. \ref{sMovingF0}, we can replace the ${SO(1,9)\over SO(1,1)\times SO(8)}$ spinor moving frame variables the by two set of spinor moving frame variables parametrizing a product of two cosets isomorphic to  8-spheres, ${\bb S}^8 \times {\bb S}^8$. We however, do not elaborate on this possibility here).

Then the non-minimal pure spinor CFT action (\ref{Snmin=}) is replaced by
\begin{eqnarray}\label{Snew=}
S_{new }= \int \left(1/2 \partial x^\mu \bar{\partial} x^\mu + p_\alpha \bar{\partial} \theta^\alpha \right) -\int \left(
  \Omega_{\bar{z}}^{\# i} {\bb D}^{= i} + \Omega_{\bar{z}}^{= i} {\bb D}^{\# i}  \right)-  \nonumber  \qquad \\
  - \int \left( \Xi^{\mp I}_{\bar{z}}D^{\pm I}_{\Lambda} +
\Xi^{(0)}_{\bar{z}}D^{(0)}_{\Lambda}+ \left(\bar{\partial} \tilde{\Lambda} - \tilde{\Lambda}
\Xi^{(0)}_{\bar{z}}
 \right)\bar{w}_{\tilde{\Lambda}} \right) +  \nonumber  \qquad \\
  +\int \left( s^{(0)}\bar{\partial}r^{(0)} +s^{+I}\bar{\partial}r^{-I} + s^{=i}\bar{\partial}r^{\# i}
 \right)
\; . \qquad
\end{eqnarray}

The spinor moving frame momentum operators ${\bb D}^{(0)}$ and ${\bb D}^{ij}$ are currents  of the $SO(1,1)$ and of the $SO(8)$  gauge symmetry of the model when acting on the spinor moving frame variable. The complete currents  of $SO(1,1)$ and of the $SO(8)$ act also on the fermionic ghosts and (in the case of $SO(1,1)$) on $SO(8,{\bb C})$ parameters. Together with the $SO(6,{\bb C})$ and $SO(2,{\bb C})$ currents they replace $J$ and $N_{ab}$  (Eqs.  (\ref{N=}), (\ref{N=bl}))
of pure spinor approach, schematically,
\begin{eqnarray}\label{Nab->}
&& N_{ab}+ {\bar N}_{ab}\mapsto \left\{\matrix{{\bb D}^{ij} + \propto (r^{\# i}s^{= j}-r^{\# j}s^{= i})\cr D_{\Lambda}^{IJ}
+ \propto (r^{-I}s^{+ J}-r^{- J}s^{+ I}), \cr  {\bb D}^{(0)} + D_\Lambda^{(0)}+ \propto (2 r^{\# i}s^{= i}-
r^{-I}s^{+ I})}\right. \; , \qquad \nonumber \\
 && J\quad \mapsto  \quad \tilde{\Lambda} {\bar w}_{\tilde{\Lambda}} + \propto ( r^{\# i}s^{= i}+
r^{-I}s^{+ I})
\; . \qquad
\end{eqnarray}

The energy--momentum tensor of the reformulated CFT (\ref{Snew=}) should read
\begin{eqnarray}\label{Tzz=}
T(z,\bar{z})= -1/2 \partial x^\mu {\partial} x^\mu - p_\alpha {\partial} \theta^\alpha  +
  \Omega_z^{\# i} {\bb D}^{= i} + \Omega_z^{= i} {\bb D}^{\# i}
 +  \Xi^{\mp I}_{{z}}D^{\pm I}_{\Lambda} +
\Xi^{(0)}_{{z}}D^{(0)}_{\Lambda}+  \nonumber  \qquad \\  + \left({\partial} \tilde{\Lambda} + \tilde{\Lambda}
\Xi^{(0)}_{{z}}
 \right)\bar{w}_{\tilde{\Lambda}}
  - s^{(0)}{\partial}r^{(0)} -s^{+I}{\partial}r^{-I} - s^{=i}{\partial}r^{\# i}
\; . \qquad
\end{eqnarray}

The BRST operator of the non-minimal pure spinor formalism is replaced by
\begin{eqnarray}\label{Qnew=}
 Q= \int \left(\Lambda^+_qD^-_q + r^{(0)} \bar{w}_{\tilde{\Lambda}}  + r^{-I}D_\Lambda^{+I} + r^{\# i}{\bb D}^{=i}  \right)\; . \qquad
\end{eqnarray}

The above presented reformulations of the elements of the nonminimal pure spinor path integral measure, Eqs.  (\ref{d11L=hOm}), (\ref{d11bL=hOm}) and (\ref{d11r=11M--}),   actually correspond to the hypothetical  quantum superstring formulation with the CFT action (\ref{Snew=}) and the BRST charge (\ref{Qnew=}). Let us stress that these are 10D Lorentz invariant; despite forms, variables and momenta, like    $\Omega_z^{= i}$, $r^{\# i}$, ${\bb D}^{=i}$ and others, carry the indices of $SO(8)$ and $SO(1,1)$ groups, these are independent gauge symmetry groups of the spinor moving frame approach, which also possesses $SO(1,9)$ symmetry acting on $x^\mu$, $\theta^\alpha$ and $p_\alpha$ (and leaving invariant  $\Omega_z^{= i}$, $r^{\# i}$, ${\bb D}^{=i}$ etc.).

Certainly, the consistency of the reformulated  theory needs to be checked.
This implies the  study  of the possible conformal anomalies of the CFT (\ref{Snew=}), as well as  of the cohomologies of the BRST operator (\ref{Qnew=}). We plan to address these problems in the nearest future.

Then, if consistency and nontriviality is proved, the next stages to develop our approach would be the construction of the b--ghost (see \cite{Oda:2007ak}) and of the vertex operators in terms of new variables, obtaining the complete expression for simplest amplitudes thus making examples of the tree and loop amplitude calculations (beginning from the ones already calculated in the pure spinor formalism, e.g. in \cite{Mafra:2011nv} and  \cite{Mafra:2012kh}).

We hope to turn to these stages  in our future publications.

{\bf Aknowledgments}.
The author is thankful to Dima Sorokin and Mario Tonin  for reading the manuscript and useful comments. A partial support from the research grants FIS2008-1980 from the
Spanish MICINN (presently MINECO) and the Basque Government Research Group Grant ITT559-10 is greatly acknowledged.

\end{document}